\newcommand{\gev}{\, {\rm GeV}}

\newcommand{\beq}{\begin{equation}}
\newcommand{\eeq}{\end{equation}}
\newcommand{\bea}{\begin{eqnarray}}
\newcommand{\eea}{\end{eqnarray}}

\newcommand{\gsim}{\lower.7ex\hbox{$\;\stackrel{\textstyle>}{\sim}\;$}}
\newcommand{\lsim}{\lower.7ex\hbox{$\;\stackrel{\textstyle<}{\sim}\;$}}

\newcommand{\trh}{T_{\rm RH}}
\newcommand{\arh}{a_{\rm RH}}

\documentclass[a4paper,11pt]{article}
\pdfoutput=1

\usepackage{jcappub}
\usepackage[utf8]{inputenc} 
\usepackage{amsfonts,amssymb,mathrsfs,amsmath,esint,bm}
\usepackage{pdflscape} 
\usepackage[capitalise]{cleveref}
\allowdisplaybreaks

\graphicspath{{./Figures/}}
\usepackage{latexsym}
\usepackage{graphicx}
\usepackage{subfigure}
\usepackage[dvipsnames]{xcolor}
\usepackage{booktabs}
\usepackage{physics}
\usepackage{tikz}
\usetikzlibrary{decorations.pathmorphing}\usetikzlibrary{positioning,calc}
\usetikzlibrary{math}
\usepackage{cleveref}
\usepackage{comment}
\usepackage[normalem]{ulem}

\hypersetup{pdfstartview=FitV,colorlinks=true,linkcolor=blue,citecolor=red,filecolor=black,urlcolor=blue}

\def\stacksymbols #1#2#3#4{\def\theguybelow{#2}
    \def\vp{\lower#3pt}
    \def\sp{\baselineskip0pt\lineskip#4pt}
    \mathrel{\mathpalette\intermediary#1}}

\def\intermediary#1#2{\vp\vbox{\sp
     \everycr={}\tabskip0pt
     \halign{$\mathsurround0pt#1\hfil##\hfil$\crcr#2\crcr
              \theguybelow\crcr}}}

\title{How Accidental was Inflation?}

\author[a,b]{\bf Ignatios Antoniadis,}
\affiliation[a]{High Energy Physics Research Unit, Faculty of Science, Chulalongkorn University, Bangkok 1030, Thailand}
\affiliation[b]{Laboratoire de Physique Th\'eorique et Hautes \'Energies
  - LPTHE \\
Sorbonne Universit\'e, CNRS, 4 Place Jussieu, 75005 Paris, France
}
\emailAdd{antoniad@lpthe.jussieu.fr}

\author[c,d]{John Ellis,}
\affiliation[c]{Theoretical Particle Physics and Cosmology Group, Department of
  Physics, King's~College~London, London WC2R 2LS, United Kingdom}
\affiliation[d]{Theoretical Physics Department, CERN, CH-1211 Geneva 23,
  Switzerland}
\emailAdd{john.ellis@cern.ch}

\author[e]{Wenqi Ke,}
\affiliation[e]{William I.~Fine Theoretical Physics Institute, School of Physics and Astronomy, University of Minnesota, Minneapolis, MN 55455, USA}
\emailAdd{wke@umn.edu} 

\author[d,f,g]{\bf \\ Dimitri V. Nanopoulos}
\affiliation[f]{Academy of Athens, Division of Natural Sciences, Athens 10679, Greece
}
\affiliation[g]{George P. and Cynthia W. Mitchell Institute for Fundamental Physics and Astronomy, Texas A\&M University, College Station, TX 77843, USA
}
\emailAdd{dimitri@physics.tamu.edu}

\author[e]{~and~Keith A. Olive,}
\emailAdd{olive@umn.edu}

\abstract{Data on the cosmic microwave background (CMB) are discriminating between different models of inflation, disfavoring simple monomial potentials whilst being consistent with models whose predictions resemble those of the Starobinsky $R + R^2$ cosmological model. However, this model may suffer from theoretical problems, 
since it requires a large initial field value, threatening the validity of the effective field theory. This is quantified by the Swampland Distance Conjecture, which predicts the appearance of a tower of light states associated with an effective ultra-violet cutoff. This could be lower than the inflation scale for cases with an extended period of inflation,  leading to an additional problem of initial conditions. 
No-scale supergravity models can reproduce the predictions of the Starobinsky model and accommodate the CMB data at the expense of fine-tuning of parameters at the level of $10^{-5}$. Here, we propose a solution to this problem based on an explicit realisation of the Starobinsky model in string theory, where this `deformation' parameter is calculable and takes a value of order of the one corresponding to the Starobinsky inflaton potential.  Within this range, there are parameter values that accommodate more easily the combination of Planck, ACT and DESI BAO data, while also restricting the range of possible inflaton field values, thereby avoiding the swampland problem and predicting that the initial conditions for inflation compatible with the CMB data are generic.}

\begin{document}
\begin{flushright}
UMN--TH--4418/25, FTPI--MINN--25/03,   \\
KCL--PH--TH/2025-09, CERN--TH--2025-076 \\
April  2025
\end{flushright}
\maketitle

\section{Introduction}

 Inflation plays a central role in the Standard Model of cosmology~\cite{reviews}. In addition to solving classic problems such as the horizon problem, the flatness problem and the isotropy problem (not to mention one of its original motivations, the magnetic monopole problem), inflation provides an explanation for the origin of the density fluctuations observable in the cosmic microwave background (CMB) that triggered the formation of astrophysical structures. 
CMB observations transformed inflation from an abstract phase of the Universe to a testable scientific theory. Indeed, observations of the spectral tilt, $n_s$, in the 2-point correlation function of the CMB anisotropy spectrum \cite{wmap,planck18}, and limits on the tensor-to-scalar ratio, $r$ \cite{rlimit,Tristram:2021tvh}, 
have the power to exclude specific models of inflation while verifying predictions of others. 

The simplest models of inflation are based on a single monomial potential, as in chaotic inflation \cite{chaotic}. However as limits on the tensor-to scalar ratio have improved, these simple models could be excluded. On the other hand, as the precisions in the measurements of $r$ and $n_s$ have improved, the Starobinsky model \cite{Staro} based on a quadratic modification of Einstein gravity, originally designed to avoid the initial singularity in Big Bang cosmology, has survived. 

In both chaotic inflation and in $R + R^2$ gravity, there is only a single parameter, the inflaton self-coupling or the coefficient of the $R^2$ term, that can be determined from experiment.
In most other models, in addition to the overall scale of inflation, there is a relative coupling that must be adjusted so that some part of the scalar potential is sufficiently flat to provide slow-roll inflation with enough $e$-folds of expansion to resolve the classic cosmological problems. 
As discussed in \cite{accident}, this necessary tuning of couplings may appear accidental, but is often required in models constructed from supergravity models.  

Fine-tuning is a recurrent issue in scalar field theories, e.g., the Higgs sector of the Standard Model electroweak sector. While the inflation scale, $M$, is generally taken to be far greater than the electroweak scale, it is still much smaller than the Planck scale: $M \sim 10^{-5} M_P$~\footnote{Here $M_P$ is the reduced Planck mass, $M_P = 1/\sqrt{8\pi G_N}$.}. Just as supersymmetry was touted as a possible solution to the Higgs fine-tuning problem, the benefits of resolving the inflationary hierarchy with supersymmetry were advertized in \cite{Cries,primordial,fluct}.

The supersymmetrisation of $R + R^2$ gravity still has a single parameter, but introduces an extra scalar degree of freedom and its supersymmetric partners, corresponding to the goldstino chiral supermultiplet and with it a tachyonic instability during inflation \cite{Cecotti, eno7}. Stabilizing the sgoldstino requires an extension of the minimal model, introducing model dependence and extra parameters. The model dependence can be eliminated by imposing nonlinear supersymmetry, which amounts to integrating out massive degrees of freedom, or equivalently imposing a constraint that eliminates the scalar partner of the goldstino (sgoldstino) \cite{Antoniadis:2014oya}. On the other hand, string theory comes to a rescue since, in the presence of the string dilaton, the sgoldstino instability is absent, at least in perturbation theory \cite{anr1, ano}.~\footnote{Of course dilaton stabilisation is required, presumably as a result of nonperturbative effects and and/or discrete fluxes.}

The construction of the Starobinsky model in the context of no-scale supergravity is a modern example of accidental inflation. The earliest example of this type is based on a Wess-Zumino form for the superpotential~\cite{eno6}, which includes quadratic and cubic couplings. The Starobinsky potential arises exactly if the ratio of these couplings is precisely $2/3\sqrt{3}$. As we discuss in more detail below, this fine-tuning must be respected to roughly 1 part in $10^5$. Furthermore, even such small changes in the couplings, have a dramatic effect on the shape of the potential, affecting the derived value of $n_s$ and the allowed range in the post-inflationary reheating temperature, $\trh$. The latter is also model-dependent, since it requires an input on how the inflaton couples to Standard Model fields, in contrast to the $R + R^2$ model,~\footnote{An elegant possibility is to identify the inflaton with the scalar partner of the right-handed neutrino (or of a linear combination of them) \cite{ENO8, snu, building}.} where this interaction arises from the universal coupling of the matter stress-tensor to the metric in the so-called Jordan frame, associated to the geometric description of the model (see for example \cite{Ema:2024sit}).

The Starobinsky model is very consistent with Planck data on $r$ and $n_s$, which is the primary reason why so much attention has been focused
on this model in the past few years. Recently the Atacama Cosmology Telescope (ACT) has released new results \cite{ACT:2025fju} that, when combined with Planck CMB data and DESI data on baryon acoustic oscillations (BAO), may indicate a somewhat larger value for $n_s$ that
puts pressure on the Starobinsky model at the $3\sigma$ level.~\footnote{We note, however, that there are subtleties in the combination of these data sets that must be addressed before firm conclusions can be drawn (D.~Scott, private communication).} As we discuss below, these results can easily be accommodated with a slightly different accidental ratio of couplings in no-scale avatars of the Starobinsky model. In other models, such as a simple model of new inflation constructed from supergravity, a comparable shift in the couplings could cause the value of $n_s$ to shift from being completely excluded ($>5\sigma$) to being in good agreement with data. 

Despite its consistency with experiment, the Starobinsky model and its supergravity generalizations are vulnerable to criticism because of the large inflaton field values ($\phi > M_P$) that they require, threatening the validity of the effective field theory and
being disfavored by the Swampland Distance Conjecture~\cite{Agrawal:2018own, Lust:2023zql}. 
In particular, the Swampland Distance Conjecture implies the appearance of a tower of light states at a scale that is exponentially suppressed by the proper distance of the field space with an exponent of order unity in Planck units \cite{Ooguri:2006in}. This tower is either of Kaluza-Klein type or a string tower associated to a low decompactification or string scale, respectively \cite{Lee:2019wij}. In both cases, there are massive spin-2 states, which should be heavier than the inflation scale by unitarity (Higuchi bound) \cite{Higuchi:1986py}, imposing a strong constraint on the inflaton initial conditions.

Interestingly, this problem can be resolved by introducing a deformation away from the Starobinsky-type potential induced by a small shift in the appropriate supergravity coupling that gives rise to a barrier in the inflaton potential, preventing the inflaton from having large field excursions. 
Such a deformation is actually calculable in string theory, as can be shown in an explicit example \cite{anr1} that was constructed in the context of free-fermionic formulation of four-dimensional heterotic superstrings \cite{Antoniadis:1986rn, Antoniadis:1987wp}. 
The quantitative analysis of this proposal in combination with the swampland constraints on the inflation initial conditions described above, is one of the main focuses of our paper.

In what follows, we examine the degree to which inflation is accidental in several well-motivated models 
based on supergravity and string theory. We derive the effects of small shifts in the couplings on the form of the potential, on the relation between $n_s$ and $\trh$ as well as its effect on limiting large field excursions. After a brief review of some basics of slow-roll inflation in Section \ref{sec:basics} we consider a sample of well-motivated models in Section \ref{examples}, including a supergravity-derived version of the Starobinsky model \cite{eno6}, a string theory-derived Starobinsky-like model \cite{anr1}, and a supergravity-based model of new inflation \cite{hrr}. We
consider the implications of the swampland distance conjecture on the Starobinsky-like models in Section \ref{swamp}. Our conclusions are given in Section \ref{summ}.

\section{Inflation basics}
\label{sec:basics}

The success of any model of inflation relies on its ability to produce density fluctuations consistent with experimental observables. In addition to providing a sufficient amount of exponential expansion to resolve the cosmological curvature and anisotropy problems, among others, these include the spectral tilt of the microwave background anisotropy spectrum and the ratio of the amplitudes of tensor and scalar perturbations. Once the inflationary potential is specified, these observables can be obtained via the slow-roll parameters, which are are determined as follows by the potential and its derivatives~\cite{reviews}:
\begin{equation} 
\epsilon \; \equiv \; \frac{1}{2} M_{P}^2 \left( \frac{V'}{V} \right)^2 ; \; \;  \eta \; \equiv \; M_{P}^2 \left( \frac{V''}{V} \right) ; \; \; \xi \equiv M_P^4 \left( \frac{V' V'''}{V^2} \right)  \, ,
\label{epsilon}
\end{equation}
where the prime denotes a derivative with respect to the canonically-normalized inflaton field, $\phi$. 

Of the inflationary observables, among the most important for constraining models is the scalar tilt, $n_s$, which can be approximated by 
\beq
n_s \;  \simeq \; 1 - 6 \epsilon_* + 2 \eta_* = \; 0.9649 \pm 0.0044 \; (68\%~{\rm CL}) \, ,
\label{ns}
\eeq
when evaluated at the pivot scale, $k_* = 0.05$~Mpc$^{-1}$. 
The numerical value in (\ref{ns}) is that determined
from Planck data~\cite{planck18} including lensing: alternative estimates including ACT data are discussed below.  
In addition to $n_s$, inflation models are constrained by the current bounds on the scalar-to-tensor ratio~\cite{rlimit,Tristram:2021tvh}:
\beq
r \; 
\simeq  \; 16 \epsilon_* < 0.036 \; (95\%~{\rm CL}) \, . \label{r}
\eeq
The overall scale of the inflationary potential is determined by the amplitude of the scalar perturbation spectrum:
\beq
A_s \; = \; \frac{V_*}{24 \pi^2 \epsilon_* M_{P}^4 } \simeq 2.1 \times 10^{-9} \, . \label{As} 
\eeq
There is currently no evidence for scale dependence (running) of $n_s$, which is of second order in the slow-roll approximation:~\footnote{Note that the value of $n_s$ in Eq.~(\ref{ns}) assumes no running. Allowing for the possibility of running, the best-fit value is $n_s = 0.9641 \pm 0.0044$ \cite{planck18}.}, 
\beq
\frac{d n_s}{d\ln k} \simeq -24 \epsilon_*^2 + + 16 \epsilon_* \eta_* -2 \xi_*^2 = -0.0045 \pm 0.0067 \, .
\eeq
Finally, the number of $e$-folds starting from the pivot scale can also be computed from the slow-roll parameters:
\begin{equation}
N_* \;\equiv\; \ln\left(\frac{a_{\rm{end}}}{a_*}\right) \; = \; \int_{t_*}^{t_{\rm{end}}} H dt \; \simeq \;  - \int^{\phi_{\rm{end}}}_{\phi_*} \frac{1}{\sqrt{2 \epsilon}} \frac{d \phi}{M_P} \, ,
\label{efolds}
\end{equation}
where $a_*$ is the value of the cosmological scale factor at the pivot scale, $a_{\rm end}$ is its value at the end of inflation (when exponential expansion ends and $\ddot{a}=0$) and $\phi_*$ and $\phi_{\rm end}$ are the corresponding values of $\phi$ 
at those two epochs. However, the total number of $e$-folds, $N_{\rm tot}$, might be much larger than $N_*$, and can be determined by replacing $\phi_* \to \phi_0$, where $\phi_0$ is the initial position of $\phi$ at some earlier time $t_0$~\footnote{We assume that $\dot\phi_0=0$, which is a suitable simplifying assumption.} and, by analogy, replacing $a_*$ (the scale factor at $t_*$) with $a_0$ (the scale factor at $t_0$).  In practice, Eq.~(\ref{efolds}) is used to determine $\phi_*$,
which is subsequently used to determine the slow-roll parameters. The number of $e$-folds, assuming no additional entropy production after reheating, is given by \cite{LiddleLeach,Martin:2010kz,EGNO5,egnov}:
\begin{equation}
\begin{aligned}
\label{eq:nstarreh}
N_{*} \; &= \; \ln \left[\frac{1}{\sqrt{3}}\left(\frac{\pi^{2}}{30}\right)^{1 / 4}\left(\frac{43}{11}\right)^{1 / 3} \frac{T_{0}}{H_{0}}\right]-\ln \left(\frac{k_{*}}{a_{0} H_{0}}\right) 
-\frac{1}{12} \ln g_{\mathrm{reh}} \\
&+\frac{1}{4} \ln \left(\frac{V_{*}^{2}}{M_{P}^{4} \rho_{\mathrm{end}}}\right) +\frac{1-3 w_{\mathrm{int}}}{12\left(1+w_{\mathrm{int}}\right)} \ln \left(\frac{\rho_{\mathrm{R}}}{\rho_{\text {end }}}\right) 
\, ,
\end{aligned}
\end{equation}
where the present Hubble parameter and photon temperature are given by $H_0 = 67.36 \, \rm{km \, s^{-1} \, Mpc^{-1}}$ \cite{planck18} and $T_0 = 2.7255 \, \rm{K}$~\cite{Fixsen:2009ug}. Here, $\rho_{\rm{end}}=\rho(\phi_{\rm end})$ and $\rho_{\rm{R}}$ are the energy density at the end of inflation and at the beginning of the radiation domination era when $w = p/\rho = 1/3$, respectively, $a_0 = 1$ is the present day scale factor, $g_{\rm{reh}} =  915/4$ is the effective number of relativistic degrees of freedom in the minimal supersymmetric standard model (MSSM) at the time of reheating, and $w_{\rm int}$ is the equation-of-state parameter averaged over the $e$-folds during reheating. Using
 $k_* \; = \; 0.05 \, \rm{Mpc}^{-1}$, the first line in (\ref{eq:nstarreh}) gives
$N_*  \simeq  61.04$. As a result,
$N_*$ depends on the reheating temperature through $\rho_{\rm R}$ and hence the inflationary observables, most notably $n_s$, do as well.  

Reheating is an essential aspect of any inflation model. In a typical single-field model, the last $N_*$ $e$-folds of inflation start when $\phi=\phi_*$ and exponential expansion stops when $\phi = \phi_{\rm end}$. 
Subsequently the inflaton begins a period of oscillations about its potential minimum, until its decay rate to light particles
becomes comparable to the expansion rate. Around that time, the radiation density becomes equal to the inflaton energy density, defining the ``moment" of reheating (though the process is not instantaneous). Thus 
\beq
\frac{g_{\rm reh} \pi^2}{30} \trh^4 = \rho_{\rm R}(a_{\rm RH}) = \rho_\phi(a_{\rm RH}) \, ,
\eeq
where $a_{\rm RH}$ is the scale factor when the inflaton and radiation densities are equal,  and
$\trh \sim (\Gamma_\phi M_P)^\frac12$,
where $\Gamma_\phi$ is the inflaton decay rate.  Depending on the specifics of the model, the decay rate may be calculable, thereby determining the specific value of the reheating temperature.
For the most part we will not assume any 
specific value for $\trh$, but simply require that $\trh \gtrsim 4$~MeV (to allow for standard Big Bang nucleosynthesis \cite{tr4})
and $\trh \lesssim (H_{\rm end} M_P)^\frac12$
where $H_{\rm end}$ is the Hubble parameter at $a_{\rm end}$.

\begin{figure}[!ht]
\centerline{\psfig{file=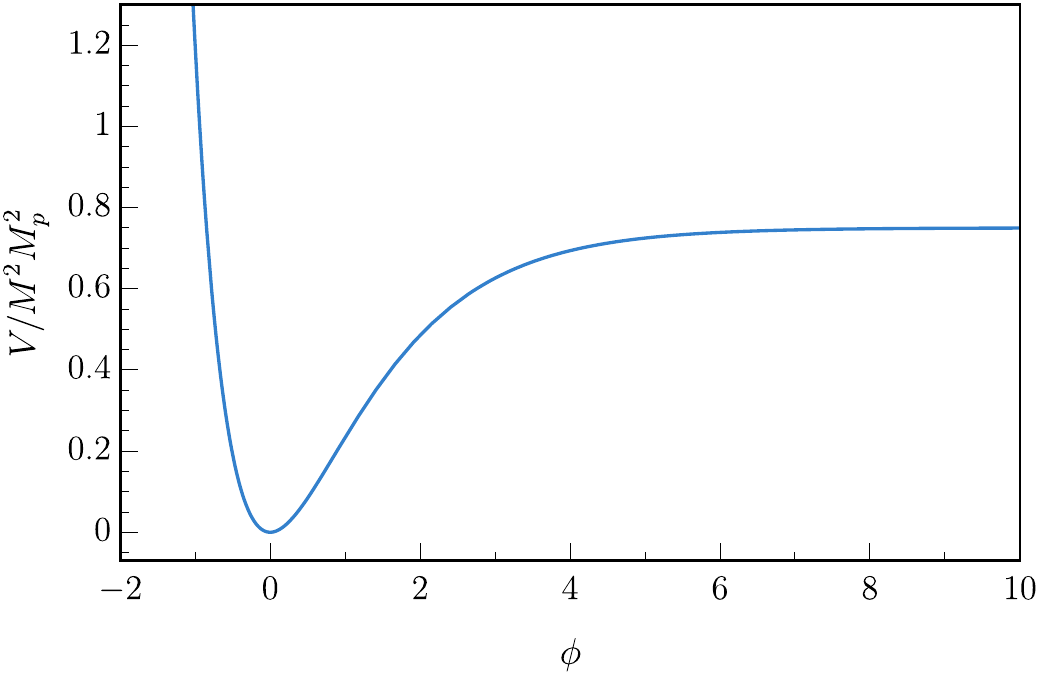,width=8.2cm}}
\caption{\it The scalar potential~(\ref{starpot}) in the Starobinsky model of inflation~\cite{Staro}. \label{fig:staro}}
\end{figure}

Before we consider examples of apparently accidental inflation, it is instructive to review the 
Starobinsky model \cite{Staro}. The model can be expressed geometrically as a modification of Einstein gravity by including a term quadratic in the scalar curvature:
\begin{equation}
{\cal A} \; = \;  \int d^4x \sqrt{-g} \left(- \frac{1}{2} R + \frac{\alpha}{2} R^2 \right) \, , 
\label{EH}
\end{equation}
where $\alpha$ is a constant\footnote{We will work in Planck units with $M_P = 1$. In certain instances to avoid confusion we will include factors of $M_P$.}.  This theory may be rewritten in terms of Einstein gravity with a massive scalar field \cite{WhittStelle,Kalara:1990ar}, 
\begin{equation}
{\cal A} \; = \; - \frac{1}{2} \int d^4x \sqrt{- {\tilde g}} \left[ {\tilde R} - \partial^\mu \phi \partial_\mu \phi + \frac{1}{4 {\alpha}} 
\left(1 - e^{-\sqrt{\frac{2}{3}} \phi }
 \right)^2 \right] \, ,
\label{FullStaro}
\end{equation}
where ${\tilde g}$ and ${\tilde R}$ are the metric and curvature scalar in the Einstein frame. 
Writing $\alpha \equiv 1/6 M^2$, we obtain the Starobinsky potential for inflation:
\beq
V(\phi) \; = \; \frac34 M^2\left(1 - e^{-\sqrt{\frac{2}{3}} \phi }
 \right)^2 \, ,
 \label{starpot}
\eeq
which is shown in Fig.~\ref{fig:staro}.
The initial value $\phi_0$ can take any value large enough to ensure sufficient inflation. As discussed above, the last $N_*$ $e$-folds begin at $\phi_*$ and end at $\phi_{\rm end}$ when oscillations about the minimum begin. 

It is straightforward to compute in the Starobinsky model the inflationary observables listed above~\cite{eno6}:
\begin{eqnarray}
A_s &  = &  \frac{3 M^2}{8\pi^2} \sinh^4 (\phi/\sqrt{6}) \, , \label{As2} \\
\epsilon & = & \frac13 \csch^2 (\phi/\sqrt{6}) e^{-\sqrt{2/3}\phi} \, ,  \label{eps}\\
\eta & = &  \frac13 \csch^2 (\phi/\sqrt{6}) \left( 2 e^{-\sqrt{2/3}\phi} -1\right) \label{eta} \, ,
\end{eqnarray} 
and $\phi_{\rm end} = \phi(a_{\rm end}) = 0.62 M_P$ \cite{EGNO5,egnov} in the Starobinsky model. As noted above, the value of $N_*$ depends on the reheating temperature.

For reheating to occur, the inflaton must decay into something that couples to Standard Model fields. For example, the simplest possibility is to consider adding just the Standard Model Lagrangian, represented here simply by the Higgs kinetic term and potential:
\begin{equation}
{\cal A} \; \supset \;  \int d^4x \sqrt{-g} \left(\partial^\mu H^* \partial _\mu H-V(H) \right) \, . 
\label{model1}
\end{equation}
to the action in Eq.~(\ref{EH}). When rewritten in the Einstein frame, this leads to 
a coupling between the inflaton and the Higgs given by 
\begin{equation}
    {\cal A}_1 \; \supset \;  \int d^4x \sqrt{- {\tilde g}}\left[-\sqrt{\frac{2}{3}}\phi \partial^\mu H^* \partial _\mu H+2\sqrt{\frac{2}{3}}\phi V(H)\right]\,, 
    \label{a1coupl}
 \end{equation}
which is dominated by the inflaton coupling to the Higgs kinetic term.  
The decay rate is then given in this simple model by
\begin{align}
	\Gamma(\phi \to HH) = \frac{N}{192\pi}\frac{M^3}{M_P^2}\,,
	\label{eq:decay_R2_A1}
\end{align}
where $N = 4$ is
the number of the real scalar degrees of freedom of the Standard Model Higgs doublet.
In the MSSM, one may take $N=98$, corresponding to the 90 real scalar degrees of freedom of squarks and sleptons plus the 8 real Higgs degrees of freedom. 
If we define the reheating temperature corresponding to $\rho_R(\arh) = \rho_\phi(\arh)$, then
\begin{align}
	\frac{g_{\rm reh} \pi^2}{30} \trh^4 = \frac{12}{25} \left(\Gamma_\phi M_P \right)^2 \, .
 \label{deftrh}
\end{align}
For $M \simeq  3.2 \times 10^{13}$~GeV (determined by Eq.~(\ref{As2})) in both the Standard Model and in the MSSM, we find~\footnote{For a more complete discussion of reheating in the Starobinsky and related models see \cite{Ema:2024sit}.}
\beq
\label{TRH}
\trh = \begin{cases}
3.2 \times 10^9~{\rm GeV} \qquad g_{\rm reh} = \frac{427}{4} \, ,
\\
1.2 \times 10^{10}{\rm GeV} \qquad g_{\rm reh} = \frac{915}{4} \,.
\end{cases}
\eeq
These values correspond to $N_* = \, 51.4 (51.8)$ and $\phi_*/M_P =5.3  \, (5.3)$ in the Standard Model (MSSM), yielding finally the predictions 
\begin{eqnarray}
n_s & = &  \begin{cases}
.9624 \qquad g_{\rm reh} = \frac{427}{4} \, ,
\\
.9628 \qquad g_{\rm reh} = \frac{915}{4} \, ,
\end{cases}
\\
r & = &  \begin{cases}
.0040 \qquad g_{\rm reh} = \frac{427}{4} \, ,
\\
.0039 \qquad g_{\rm reh} = \frac{915}{4} \, .
\end{cases}
\end{eqnarray}
These theoretical values can be compared with the experimental values in Eqs.~(\ref{ns}) and (\ref{r}), and are in excellent agreement. 

We consider below formulations \cite{eno6} of Starobinsky-like models based on no-scale supergravity and string theory. In those models the reheating temperature is significantly more model-dependent and a wider range of values of $n_s$ and $r$ are possible, particularly if does not require the 
accidental choice of couplings leading to the Starobinsky potential.

Results from the ACT have recently been released \cite{ACT:2025fju}. A fit to the ACT data alone
yields $n_s = 0.9666 \pm 0.0077$, which is
consistent with the Planck measurement and in excellent agreement with the Starobinsky model. 
However, due to the opposite correlation between $n_s$ and $\Omega_b h^2$ in the two data sets, a fit to the combination of Planck and ACT data
results in a higher value, $n_s = 0.9709 \pm 0.0038$, which is only marginally consistent with the Starobinsky model. 
The result is little changed when lensing is included, becoming $n_s = 0.9713 \pm 0.0037$. However, when BAO data from DESI~\cite{DESI:2024mwx} are included the result for $n_s$ is further increased to $0.9743 \pm 0.0034$, which is in tension with the Starobinsky model by more than $3\sigma$.

Playing Devil's advocate, in the following we compare the impact on models of this combined Planck-ACT-DESI result as well as the result (\ref{ns}) from Planck data alone.

\section{Examples}
\label{examples}

Most models of inflation have, to some extent, two fine-tunings~\footnote{A notable exception to this is chaotic inflation~\cite{chaotic} with a monomial scalar potential.}. One fine-tuning involves the scale of inflation: $M \ll M_P$ in order to fit the CMB normalization parameterized by $A_s$, and
the other is needed to obtain a flat potential. Here we are primarily focused on this latter fine-tuning. We consider in this Section some well-motivated no-scale supergravity models that lead to Starobinsky-like inflation and a simple supergravity model of ``new" inflation. In all of the examples, some degree of relative fine-tuning between the couplings is necessary, in order that some part of the potential in field space be relatively flat.

\subsection{No-Scale Starobinsky-like Models}

In the Starobinsky model as derived from Eq.~(\ref{EH}), the flat potential shown in Fig.~\ref{fig:staro} arises without the need for fine-tuning, as the only additional parameter in the model determines the inflationary mass scale and is fixed by the CMB normalization $M = \sqrt{6\alpha} \simeq 3.2 \times 10^{13}$~GeV. The same potential
can arise quite simply~\cite{eno6,building} in no-scale supergravity~\cite{no-scale1}.
The minimal no-scale model appropriate for  inflation involves two complex fields: the volume modulus $T$, and a matter-like field $\chi$. They parametrize a non-compact
SU(2,1)/SU(2)$\times$U(1) coset space, whose field-space geometry is described by the K\"ahler potential
\beq
\label{v0}
K \; = \; -3 \ln (T + T^* - |\chi|^2/3) \, .
\eeq
The Starobinsky potential arises from this theory if one postulates a Wess-Zumino-like superpotential~\cite{eno6}
\begin{equation}
W(\phi) = {M}\left( \frac12 \chi^2 - \frac{\lambda}{3\sqrt{3}}\chi^3 \right) \, ,
\label{wi}
\end{equation}
and assumes that a) the modulus is dynamically fixed so that $\langle T \rangle = \frac12$,
and b) $\lambda = 1$.  The field $\chi$ is not canonical but, once $T$ is fixed, one can define the canonical field
\beq
\phi \; \equiv \; \sqrt{3} \tanh^{-1} \left( \frac{\chi}{\sqrt{3}} \right) \, ,
\label{can1}
\eeq
and obtain the Starobinsky potential $V(\phi)$.
The assumption of a specific value of $\lambda$ is the main focus of our analysis of this model and the analogous assumptions in other related models. 

\begin{figure}[ht!]\hspace{15mm}
\includegraphics[width=0.75\textwidth]{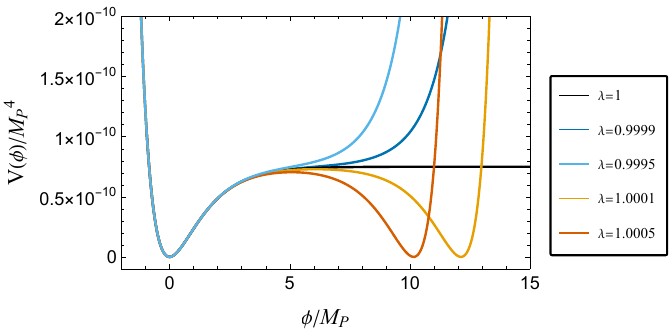}
\caption{\it The inflaton potential constructed from Eqs.~(\ref{v0}) and (\ref{wi}) for selected values of $\lambda$, assuming $M=10^{-5}M_P$.}
\label{eno6lambda}
\end{figure}

The scalar potential for the canonical field $\phi$ is easily derived from the superpotential (\ref{wi}):
\beq
V(\phi) = 3 M^2 \sinh ^2(\phi/\sqrt{6}) \left(\cosh(\phi/\sqrt{6})-\lambda \sinh(\phi/\sqrt{6})\right)^2
\eeq
which reduces to Eq.~(\ref{starpot})
for $\lambda = 1$. 
The scalar potential is shown in Fig.~\ref{eno6lambda} for several values of $\lambda$. As one can see, for values of $\lambda$ slightly less than one, there is a potential wall preventing the inflaton from attaining large field values (we return to this in Section~\ref{swamp} in relation to the swampland conjecture), whereas for $\lambda$ slightly larger than one, there is a second minimum and inflation requires an initial value of $\phi$ close to the local maximum of the potential. In either case, it is clear that successful inflation can occur in this model only if $\lambda$ is ``accidentally" close to one. 

Although the scalar potential for the inflaton in the Starobinsky $R + R^2$ model
and its no-scale avatar are identical when $\lambda = 1$, reheating is not, in general~\cite{Ema:2024sit}. As noted above, reheating in the Starobinsky model leads to a relatively high reheating temperature through the conformal coupling of the inflaton to matter in the Einstein frame. In the no-scale case with superpotential (\ref{wi}), if there is no additional superpotential coupling between the inflaton and SM fields, there are no decay terms for the inflaton to scalars or fermions \cite{EKOTY,EGNO4}. Alternatively, coupling the inflaton to $H L$ (as in models where the inflaton is a right-handed sneutrino \cite{ENO8,snu}) or through the gauge kinetic function \cite{EKOTY,EGNO4,klor,gkkmov} will lead to reheating at some model-dependent scale.
Unlike the case of the Starobinsky model, the scalar tilt will depend on this unknown reheating temperature, which we treat as a free parameter. 

\begin{figure}[ht!]\hspace{18mm}
\includegraphics[width=0.75\textwidth]{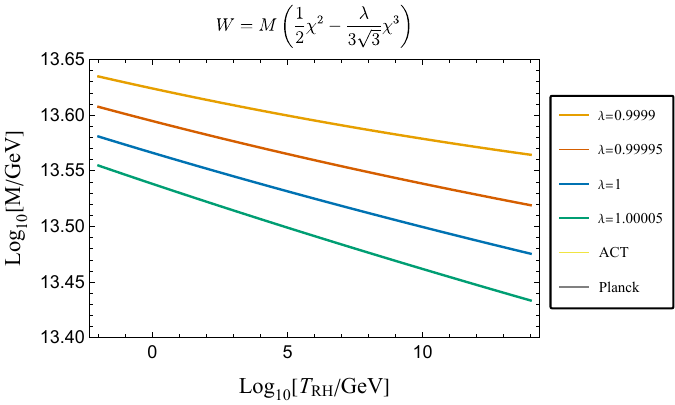}
\caption{\it The relation between $M$ and $\trh$ obtained by varying $\lambda$ in Eq.~(\ref{wi}).}
\label{Mvtrh}
\end{figure}

As discussed above, there is a logarithmic dependence on $\trh$ in the calculation of $N_*$, which in turn determines $\phi_*$ and consequently $\epsilon_*$ and $\eta_*$, and ultimately $n_s$ and $r$.
The inflationary scale, $M$, is determined by Eq.~(\ref{As}) and is not very sensitive to either $\lambda$ or $\trh$, as seen in  Fig.~\ref{Mvtrh}. In the Starobinsky model, $M$ is also the inflaton mass.

\begin{figure}[ht!]\hspace{15mm}
\includegraphics[width=0.75\textwidth]{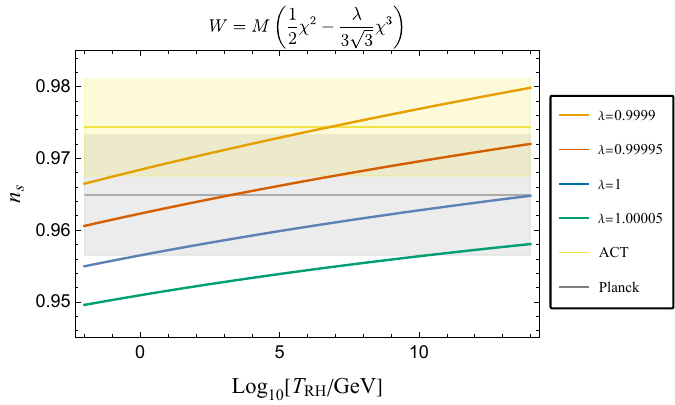}
\caption{\it Predictions for $n_s$ as a function of $\trh$ for different choices of $\lambda$ in the no-scale supergravity model with a Wess-Zumino superpotential. The grey and yellow shaded bands show $\pm 2\sigma$ ranges around the central values of the $n_s$ measurements by Planck and the Planck+ACT+DESI BAO combination, respectively.}
\label{nsvtrh}
\end{figure}

The experimental precision in $n_s$ provides significant sensitivity to both $\lambda$ and $\trh$. The value of $n_s$ as a function of $\trh$ is shown for several values of $\lambda$ in Fig.~\ref{nsvtrh}, including the choice $\lambda = 1$ that corresponds to the Starobinsky potential (blue line). The $2\sigma$ range of the Planck measurement is shown by the shaded grey band centred on a central solid line. We see that the Starobinsky model is within the $2\sigma$ range of the Planck result for all reheating temperatures from $\sim 1 - 10^{14}$~GeV, which is one of the reasons so much attention has been focused on the Starobinsky model.  In contrast, we see that, for $\lambda = 1.00005$, the model is consistent with the Planck data at the $2\sigma$ level only for $\trh > 10^{11}$~GeV. On the other hand, for $\lambda = 0.9999$, the Planck data require $\trh \lesssim 10^5$~GeV at the $2\sigma$ level and, for $\lambda < 0.99984$, $n_s$ is outside the Planck $2\sigma$ range for all reheating temperatures above 4 MeV. 

These conclusions are altered if we take into account the recent results from ACT when combined with Planck and DESI BAO data~\cite{ACT:2025fju}. Taking $n_s = 0.9675 - 0.9811$ as the $2\sigma$ range, we would conclude that the Starobinsky/$\lambda = 1$ no-scale supergravity model is always outside the $2\sigma$ range. Furthermore, the model with $\lambda = 0.9999$ lies fully within the Planck+ACT+BAO result. This suggests a need to understand better the origin of the ratio of the quadratic and cubic superpotential couplings in Eq.~(\ref{wi}). We return to this point in Subsection \ref{ANR}. 

In view of the extent to which the shape of the potential is altered for $\lambda = 0.9999$ (see Fig.~\ref{eno6lambda}), it is important to check that the evolution of the inflaton still leads to successful inflation without an additional need to  fine-tune its initial conditions, even though the shape of the potential between $\phi_*$ and $\phi_{\rm end}$ provides a good value for $n_s$.

\begin{figure}[!ht]
\includegraphics[width=\textwidth]{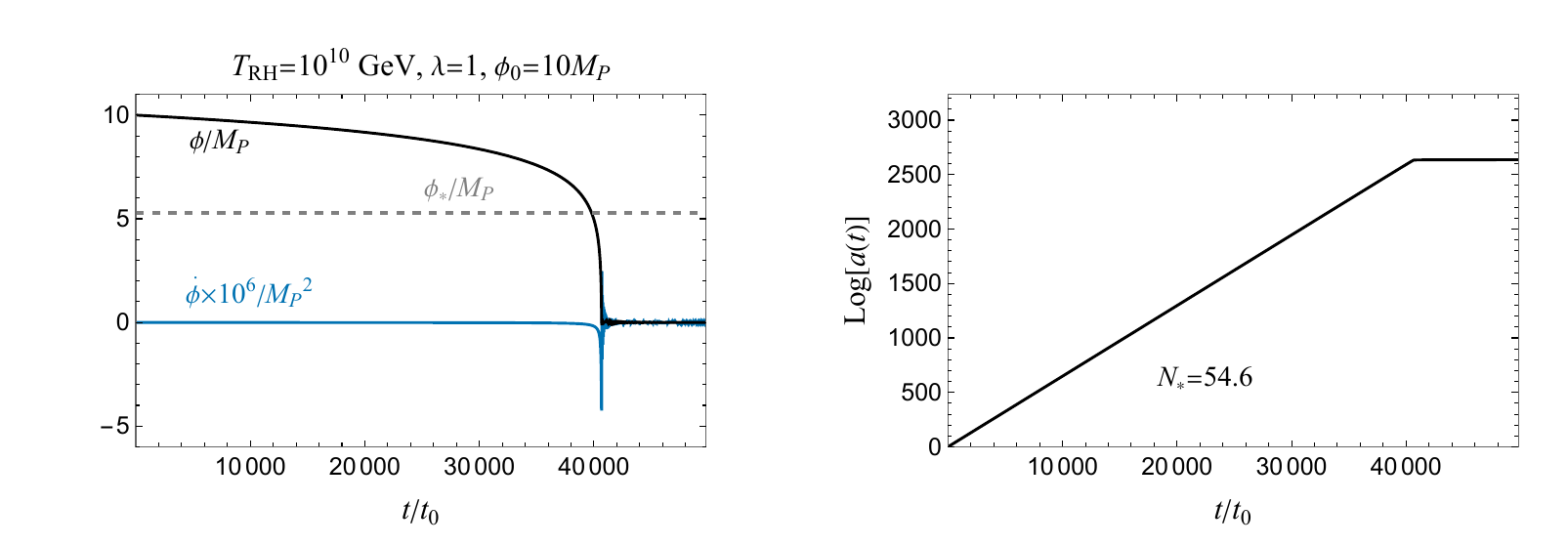}
\caption{\it The evolution of the inflaton as a function of time (left) and the total number of $e$-folds of inflation (right) for $\lambda =1$, $\trh = 10^{10}$~GeV, $\phi_0 = 10M_P$, and ${\dot{\phi}}_0 = 0$. In the right panel, we also give the value of $N_*$, the number of $e$-folds between the pivot scale and the end of inflation.}
\label{vlam}
\end{figure}

We show in Fig.~\ref{vlam} the evolution of the inflaton as a function of time in the left panel and total number of $e$-folds, in the right panel, attained for an initial value $\phi_0 = 10 M_P$ and ${\dot{\phi}}_0=0$, for the Starobinsky model
with $\lambda = 1$ and $\trh = 10^{10}$~GeV. The horizontal dashed line in the left panel corresponds to the value of $\phi_*$. The evolution of the inflaton field value is shown by the solid black line, and the evolution of $\dot \phi$ is shown by the solid blue line.
We see that the inflaton evolves slowly at first (slow roll)
leading to the quasi-linear growth in $\log a$ (exponential expansion) shown in the right panel. Starting at the somewhat arbitrary value of $\phi_0 = 10 M_P$, we see that the total number of $e$-folds is of order 2500, while $N_* = 54.6$ in this case. The total number of $e$-folds is very sensitive to the choice of $\phi_0$.  

\begin{figure}[ht!]
\includegraphics[width=\textwidth]{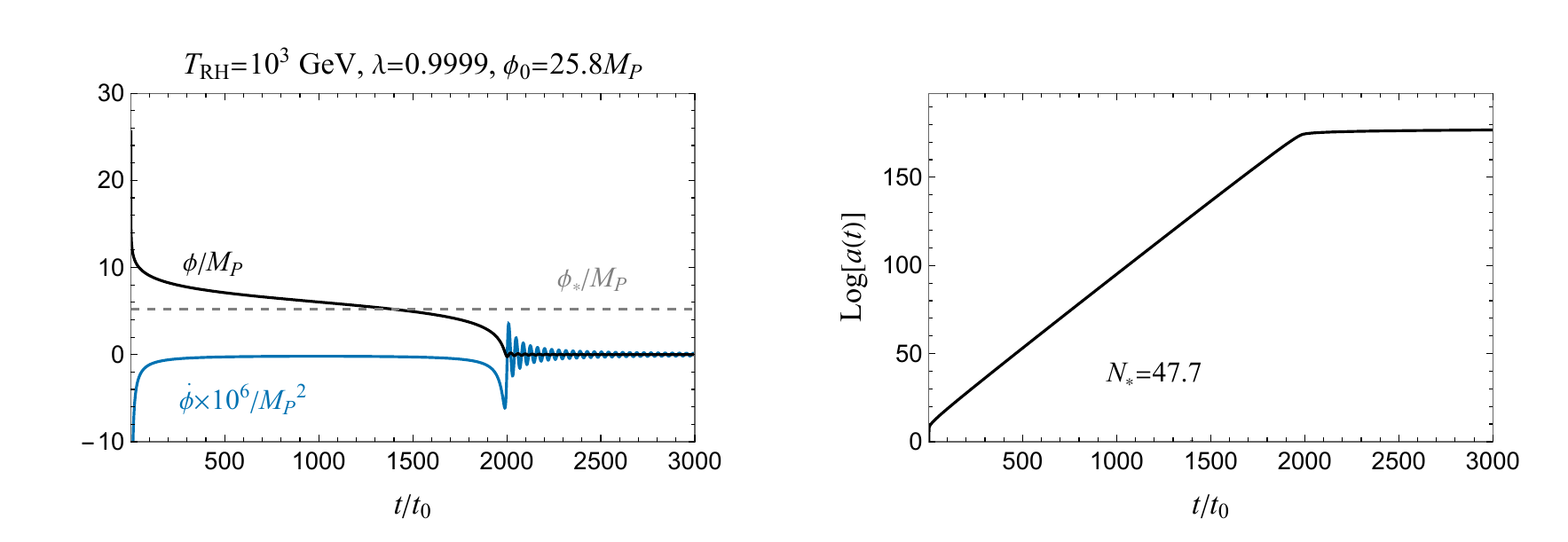}
\includegraphics[width=\textwidth]{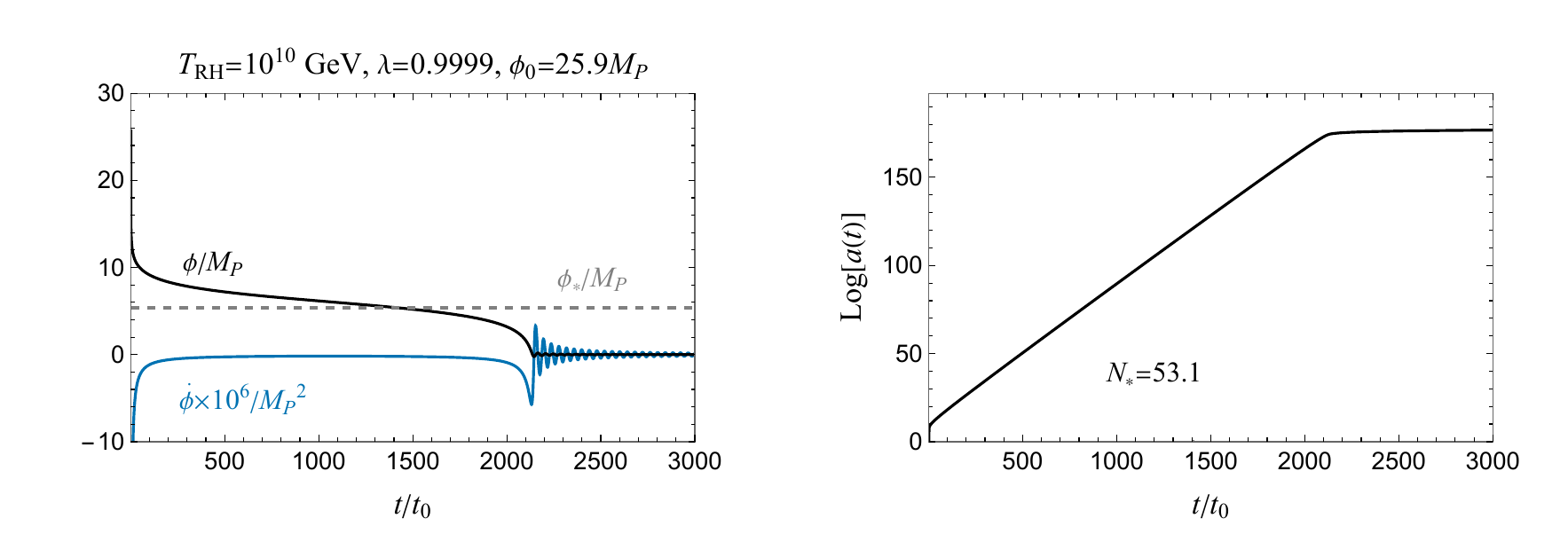}
\caption{\it The evolution of the inflaton as a function of time (left) and the total number of $e$-folds of inflation (right) for for $\lambda =0.9999$, $\phi_0 \simeq 26 M_P$, and ${\dot{\phi}}_0 = 0$. We assume $\trh = 10^3$~GeV in the upper panels and $\trh = 10^{10}$~GeV in the lower panels. In the right panels, we also give the values of $N_*$ for the number of $e$-folds between the pivot scale and the end of inflation.}
\label{vlam1}
\end{figure}

Next, we show in Fig.~\ref{vlam1} the case $\lambda = 0.9999$ and $\trh = 10^3$~GeV (top panels) and $\trh = 10^{10}$~GeV (lower panels). Although the potential becomes very steep for $\phi > \phi_*$, we see that successful inflation still occurs. Taking the maximal value $\phi_0 \simeq 26 M_P$, for which $V(\phi_0) = M_P^4$, we see that, whilst the inflaton begins rolling very rapidly, it slows quickly due to Hubble friction before it rolls to its minimum and begins oscillations about the minimum. 
Whilst the total number of $e$-folds is significantly less than in the model with $\lambda = 1$ and $\phi_0=10 M_P$, there are still over 170 $e$-folds, more than enough to solve the cosmological problems that inflation sets out to solve. As one can see, there is very little dependence on the evolution of the fields and the reheating temperature, which affect only slightly the values of $N_*, \phi_*$ and $n_s$.

\begin{figure}[ht!]
\includegraphics[width=\textwidth]{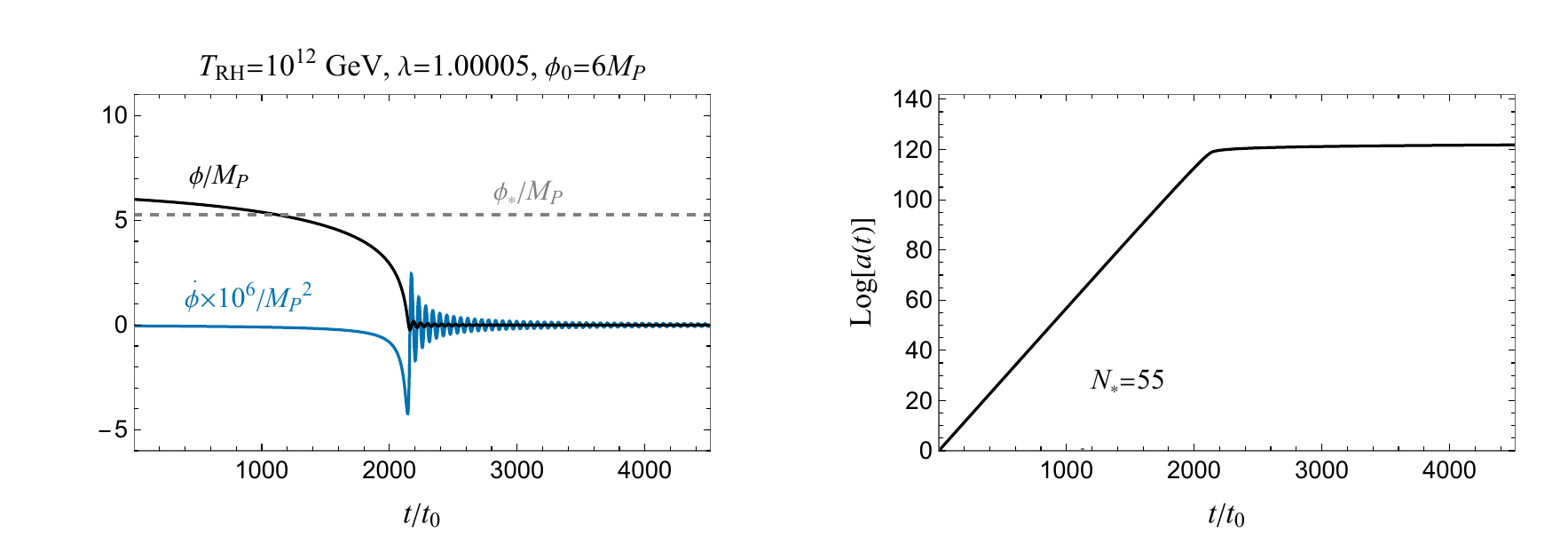}
\caption{\it The evolution of the inflaton as a function of time (left) and the total number of $e$-folds of inflation (right) for $\lambda =1.00005$, $\trh = 10^{12}$\gev, $\phi_0 = 6M_P$, and ${\dot{\phi}}_0 = 0$.}
\label{vlam2}
\end{figure}

For completeness, we show in Fig.~\ref{vlam2} an example of the evolution when $\lambda = 1.00005$. In this case $\phi_0$ must be chosen near the local maximum of the potential around $6 M_P$, and the model becomes reminiscent of hilltop inflation.

The superpotential (\ref{wi}) is not the only choice that yields a Starobinsky-like model \cite{eno7,enov1}. Enlarging the class of no-scale models is most easily done by first performing a field redefinition to a symmetric basis:
\begin{equation}
T \; = \; \frac{1}{2} \left( \frac{1 - {y_2}/{\sqrt{3}}}{1 + {y_2}/{\sqrt{3}}} \right)\, ; \qquad \;
\chi \; = \; \left(\frac{y_1}{1 + {y_2}/{\sqrt{3}}} \right) \, ,
\label{Tphirewrite}
\end{equation}
in which the K\"ahler potential can be expressed as 
\begin{equation}
K \; = \; - 3 \ln \left(1 - \frac{|y_1|^2 + |y_2|^2}{3} \right) \, .
\label{K21symm}
\end{equation}
The superpotential (\ref{wi}) then becomes
\beq
M \left[ \frac{y_1^2}{2}\left(1+  \frac{y_2}{\sqrt{3}} \right) - \lambda \frac{y_2^3}{3\sqrt{3}} \right] \, ,
\label{wsb1}
\eeq
and the Starobinsky potential is realized when $\langle y_2 \rangle = 0$, $\lambda = 1$
and one may make a redefinition similar to Eq.~(\ref{can1}) to a canonical field $\phi$. 

It is possible to construct related superpotentials consistent with the SU(2,1)/SU(2)$\times$U(1) symmetry~\cite{enov1}.
One example is
\beq
W = M y_1 y_2 \left(1+ \lambda \frac{y_2}{\sqrt{3}} \right) \, ,
\label{chargestaro}
\eeq
which leads to exactly the same results as the model discussed previously upon varying $\lambda$.
Choosing $\lambda = 1$ and applying the inverse of the transformation (\ref{Tphirewrite}), 
we obtain another 
well-studied example \cite{Cecotti}:
\begin{equation}
W(\phi) = \sqrt{3}{M} \chi \left( T-\frac12 \right) \, ,
\label{wii}
\end{equation}
which also gives the Starobinsky model in the appropriate limiting case $\langle \chi \rangle = 0$ ($ \langle y_1 \rangle = 0$)
\cite{FeKR,EGNO2,others,EGNO3}.
In this case, it is the canonically redefined $T$-field that acts as the inflaton:
\beq
\phi = \sqrt{\frac32} \ln (2 T) \, .
\label{canphi}
\eeq
Actually, this model reproduces precisely the supersymmetrisation of the Starobinsky bosonic action $R+R^2$ which can be linearised in terms of two chiral superfields $T$ and $\chi$ coupled to ordinary $N=1$ supergravity \cite{Cecotti}; $T$ contains the inflaton and $\chi$ contains the goldstino as can be seen by the linear term in the superpotential \eqref{wii}. However, as mentioned in the introduction the scalar partner of the goldstino (sgoldstino) is tachyonic during inflation (large $\text{Re}T$) and a modification of the model is needed to stabilize this direction (as is the case for generic no-scale Starobinsky-like models), that we don't address in this work. The effective supergravity upon stabilization can be described as a non linear supersymmetric constraint for the goldstino superfield $\chi^2=0$ \cite{Antoniadis:2014oya}, or equivalently for the chiral curvature superfield $({\cal R}-\mu)^2=0$ with $\mu$ constant \cite{Dudas:2015eha, Antoniadis:2015ala}, that eliminates the (massive) sgoldstino from the low energy spectrum.

As in the case of $R+R^2$ gravity, the coupling of the inflaton to matter can be introduced in a universal way by adding the MSSM Lagrangian in the geometric Jordan frame, as described in Section 2. The reheating temperature is then calculable, given in \eqref{TRH}. Although this model has again one parameter, a deformation $\lambda\ne 1$ can be introduced in the superpotential \eqref{chargestaro}, or equivalently in the K\"ahler potential \eqref{K21symm} upon appropriate field rescaling. One can then easily see that the divergence in the scalar potential corresponds to the boundary of the K\"ahler cone which restricts the allowed field space.

As other variants do not provide additional insight, we do not discuss them further, other than noting that in principle they display the same ``accidental" feature as the Wess-Zumino model (\ref{wi}).

\subsection{ANR model}
\label{ANR}

As described in the Introduction, the ANR model is the inflaton sector of a four-dimensional heterotic string ground state with $N=1$ supersymmetry, whose low-energy spectrum includes (i) an observable sector with gauge group the flipped $SU(5)\times U(1)$ \cite{Barr,DKN,Antoniadis:1989zy,anr2}, three generations of quarks and leptons and a pair of GUT Higgses that can break the group to the Standard model as well as 3 pairs of electroweak Higgs doublets; (ii) a hidden sector with gauge group $SO(10)\times SO(6)$ and non-chiral matter representations; (iii) an abelian $U(1)^4$ factor, one linear combination of which is anomalous leading to a constant contribution in its D-term which is a small parameter $\zeta$ proportional to the one-loop anomaly. As a result, supersymmetric conditions of F- and D-flatness lead to a set of vacuum expectation values (VEVs), 
whose magnitude is set by $\zeta$, that break in particular all $U(1)$ symmetries and generate a superpotential calculable in the $\alpha'$-expansion, or equivalently in powers of $\zeta$. 

A consistent solution of the F- and D-flatness equations, breaking the flipped $SU(5)$ to the Standard model and generating a hierarchical pattern of Yukawa couplings for quarks and leptons, leads to a very interesting phenomenology that has been studied in the past (see \cite{anr2} and references therein). 
Inflation arises during the unbroken gauge group phase, driven by an inflationary sector of two chiral superfields that generate a supersymmetric version of $R + R^2$ no-scale supergravity. The inflaton superfield is a chiral multiplet that mixes with those of the right-handed neutrinos while the sgoldstino is stable during inflation due to the presence of the string dilaton. 

The $N=1$ effective supergravity is calculable to all orders in the inverse string tension $\alpha'$-expansion and the superpotential is generated at 6th and 8th order in the perturbative expansion away from the free-fermionic point~\cite{anr1}, driven by the cancellation of a $U(1)$ anomaly, that fixes the inflation scale within the energy range required by observations. The superpotential is a sum of two terms with a relative coefficient which is very close to the value corresponding to the Starobinsky-type potential but not identical. Thus, the small deformation needed for the barrier of the inflaton initial conditions described above is calculable and depends on the details of the model. 

The K\"ahler potential in this model can be expressed as\footnote{Here, we ignore the role of the dilaton.} \cite{anr1}
\beq
\label{charge4}
K_\text{string}=-2\ln\left(1-|y|^2\right)-2\ln\left(1-\frac12|z|^2\right)\,,
\eeq
with the same superpotential as in (\ref{chargestaro}) with the replacement $y_1 \to \sqrt{3}z$ and $y_2 \to -\sqrt{3}y$:
\beq\label{Wstring}
W_\text{string}=Mz(y-\lambda y^2)\,,
\eeq
where the scale $M$ is of order $\zeta^5 M_s$ with $M_s$ the string scale, providing the required hierarchy for the inflation scale $\sim {\cal O}(10^{13})$ GeV.
Moreover, the sgoldstino is fixed at the origin $z=0$ when taking into account the dilaton contribution.
After a further field redefinition, we define the canonical field
\beq
\phi \; \equiv \; 2 \tanh^{-1} y
\label{can2}
\eeq
and obtain the following scalar potential for $\phi$:
\beq
V(\phi) = \frac14 M^2 \left(\sinh (\phi) -2 \lambda \sinh^2(\phi/2)\right)^2 \, ,
\eeq
which reduces to the Starobinsky-like potential \cite{anr1,ano}
\beq
V = \frac{M^2}{4}(1-e^{-\phi})^2 \, ,
\label{ANRpot}
\eeq
for $\lambda = 1$, albeit with a different normalization of the inflaton field in the exponential. For $\lambda \ne 1$, the potential is qualitatively similar to that shown in Fig.~\ref{eno6lambda}. Actually $\lambda={\cal O}(1)$ is {\em not accidental}, because it is given by the ratio of two terms appearing at the same order $\zeta^5$ in the $\alpha'$-expansion, when the $\zeta$-dependence of the VEVs is taking into account. Its value is calculable in principle, but depends on the details of the model defined by the set of VEVs that solve the F- and D-flatness equations~\cite{anr1}. A naive estimate would give a value less than 1, since the cubic term in the superpotential \eqref{Wstring} is generated at higher order when the $\zeta$-dependence of the VEVs is not taken into account, although a more precise calculation is required and its model dependence should be investigated.  In general, its expression is given by:
\beq
\lambda\simeq \zeta \frac{C_8}{C_6}
\frac{\langle{D_1}{\cdot}{D_4}\rangle}{ \langle{D_1}{\cdot}{D_5}\rangle}\,,
\eeq
where $C_6$ and $C_8$ are numerical coefficients multiplying the 6th and 8th order operators while $D_i$'s are gauge non-singlet fields from the hidden sector satisfying $\langle{D_1}{\cdot}{D_4}\rangle\sim\zeta^2$ and $\langle{D_1}{\cdot}{D_5}\rangle\sim\zeta^3$.

\begin{figure}[ht!]\hspace{18mm}
\includegraphics[width=0.75\textwidth]{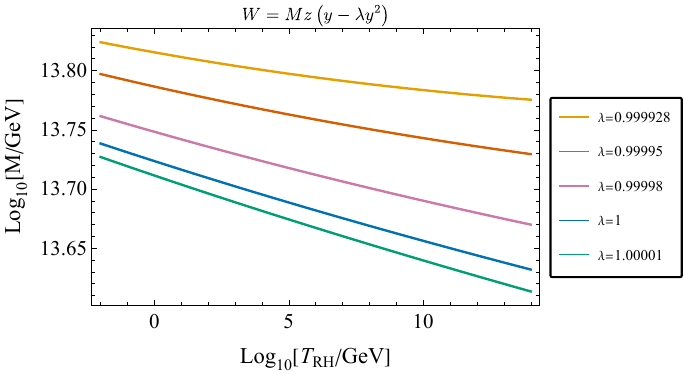}
\caption{\it Values of $M$ and $\trh$ obtained by varying $\lambda$ in the ANR model.}
\label{MANR}
\end{figure}

\begin{figure}[ht!]\hspace{15mm}
\includegraphics[width=0.75\textwidth]{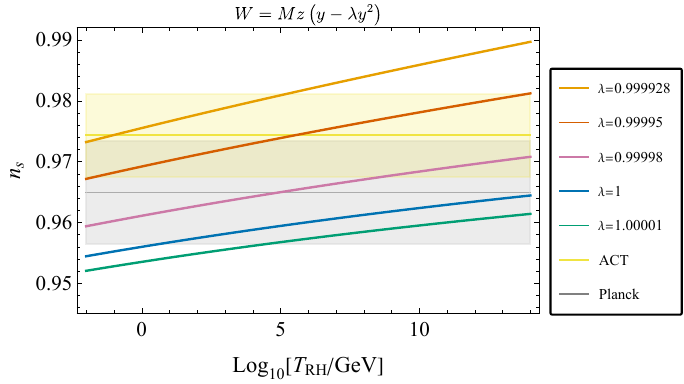}
\caption{\it Predictions for $n_s$ as a function of $\trh$ for different choices of $\lambda$ in the ANR
model. The gray and yellow shaded bands represent the $n_s$ measurements by Planck and ACT respectively, up to $\pm 2\sigma$ deviations.}
\label{nsvtrhanr}
\end{figure}

As in the model presented in the previous Section, the normalization of the potential, $M$, is not very sensitive to either  $\trh$ or $\lambda$, as shown in Fig.~\ref{MANR}.  (In this model the inflaton mass is given by $M/\sqrt{2}$.) Once again, we also see a relatively strong dependence of $n_s$ on $\trh$, and even more so on $\lambda$. While the base model with $\lambda = 1$ is consistent with the Planck result for $\trh \gtrsim 10$~GeV, it is not consistent at the $2\sigma$ level with the Planck + ACT + BAO combination unless $\lambda \simeq 0.99995$. This behavior is shown in Fig.~\ref{nsvtrhanr}.

\subsection{New Inflation in Minimal Supergravity}

Up to now, we have concentrated on models derived from no-scale supergravity, leading to inflaton potentials similar to the Starobinsky model. Alternatively, it is possible to construct models of inflation in the context of minimal $N=1$ supergravity \cite{nost,hrr,kyy}. In this case the K\"ahler potential is simple:
\beq
K = \phi^*_i \phi_i
\eeq
for all chiral superfields $\phi^i$.
Constructions in minimal supergravity are generally problematic as the scalar potential is in general not positive semi-definite and AdS
minima are common occurrences.
Another difficulty has to do with the flatness of the potential needed for inflation. Since $V \propto e^K = e^{ \phi^*_i \phi_i}$, mass terms of order the Hubble scale are generated for all fields including the inflaton. This is known as the $\eta$-problem \cite{eta}. Specific models can avoid this problem, which is generally avoided in no-scale models \cite{GMO}.  

\begin{figure}[ht!] \hspace{16mm}
\includegraphics[width=0.55\textwidth]{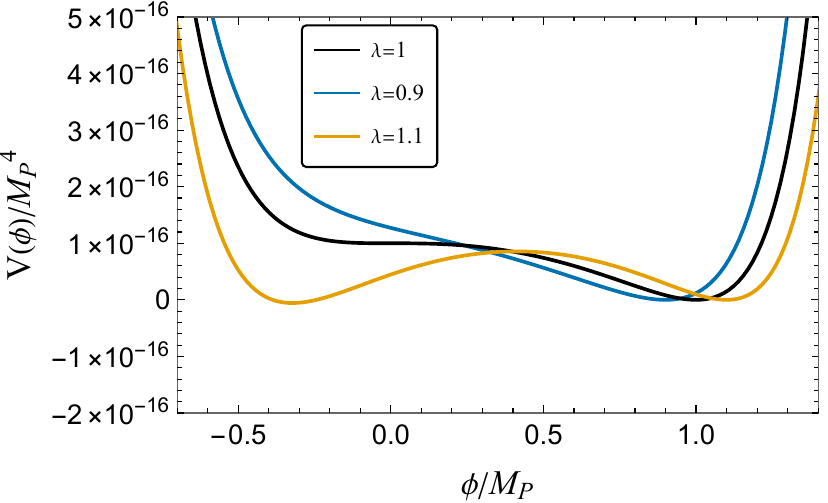} 
\caption{\it  The inflaton potential in the HRR model for selected values of $\lambda$, assuming $M=10^{-8} \gev$. }
\label{newinfpot} 
\end{figure}

\begin{figure}[ht!] \hspace{20mm}
\includegraphics[width=.55\textwidth]{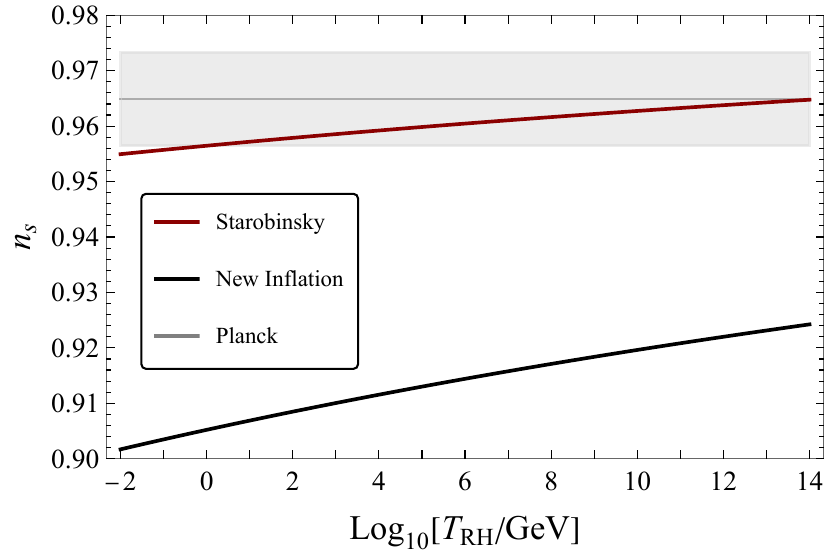}
\caption{\it Predictions for $n_s$ as a function of $\trh$ for $\lambda = 1$ in the HRR model, compared with the Starobinsky model. The horizontal solid and dashed grey lines represent the central value and the $2\sigma$ range of the $n_s$ measurements by Planck.}
\label{nsvtrhrr}
\end{figure}

One of the simplest examples of an inflation model in minimal supergravity is described by the superpotential \cite{hrr}, which we term the HRR model:
\beq
W = M (\phi - \lambda)^2 \, ,
\label{hrr}
\eeq
which leads to 
\begin{eqnarray}
V &  = & M^2 e^{|\phi|^2} \left[ -3 \left| \lambda- \phi \right|^4 + \left| \lambda- \phi \right|^2 \left|2-\phi^* (\lambda - \phi) \right|^2 \right] \, .
\end{eqnarray}
This can be expanded about the origin in the real direction $\phi = \phi^*$ to give 
\beq
V \simeq M^2\left(4 \lambda^2 - 3\lambda^4 +8 (\lambda^3-\lambda) \phi - 2(\lambda^4+\lambda^2-2)\phi^2 +(4 \lambda^3-8\lambda) \phi^3 (-\frac12 \lambda^4 + 2\lambda^2 +5) \phi^4 + \dots  \right)
\eeq
or
\beq
V \simeq M^2 \left(1 - 4 \phi^3 + \frac{13}{2} \phi^4 + \dots \right) \, ,
\eeq
for $\lambda=1$. This potential 
is shown in Fig.~\ref{newinfpot} for three values of $\lambda$. 
This is an example of new inflation~\cite{new} driven by the cubic term,
and the mass scale $M \sim 1.3\times 10^{-8} M_P$ is determined by the normalization of the CMB fluctuation spectrum (\ref{As}). This potential has a minimum at $\phi = \lambda$ with $V = 0$ for $\lambda \le 1$. For $\lambda > 1$, there is an additional minimum which is an AdS minimum when $\lambda \gtrsim 1.1$. However, as we see below, models with $\lambda > 1$ give a very small value of $n_s$. In Fig.~\ref{newinfpot}
we show exaggerated values of $\lambda \ne 1$ to make the differences in $V$ apparent. 
Although the small variations of $\lambda$ that are necessary to obtain acceptable 
values of $n_s$ would look identical in the plot, the shape of the potential is sufficiently altered (for example, the inflection point is only present for $\lambda = 1$) to make significant changes in $n_s$, as seen in Fig.~\ref{nsvtrhrr}. 
As shown there, the minimal HRR model with $\lambda = 1$ has been ruled out, since it predicts $n_s = 0.92$, which is far below the Planck lower range. In this figure we compare the HRR directly with the Starobinsky model in relation to the Planck range in $n_s$. 

\begin{figure}[ht!]\hspace{15mm}
\includegraphics[width=.75\textwidth]{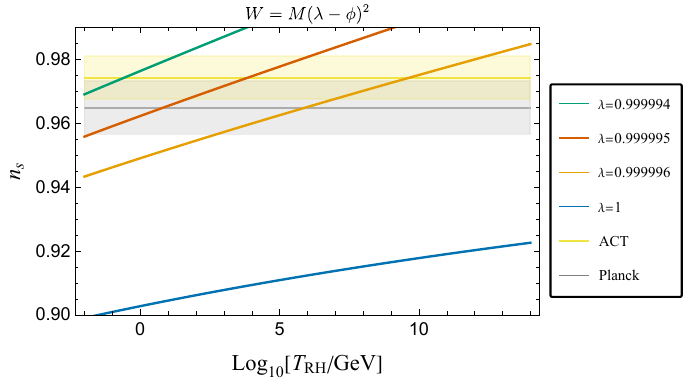}
\caption{\it Predictions for $n_s$ as functions of $\trh$ for different choices of $\lambda$ in the HRR model.  
The
gray and yellow shaded bands represent the $n_s$ measurements by Planck and Planck+ACT+DESI, respectively.}
\label{vlamhrr}
\end{figure}

However, as seen in Fig.~\ref{vlamhrr} the predicted value of $n_s$ changes rapidly for $\lambda < 1$, 
rendering the model acceptable.~\footnote{Moreover, the $\lambda = 1$ HRR model can be saved if a curvaton is present in the theory \cite{Choi:2024ruu}. }
In the figure, we show three values of $\lambda = 0.999994-0.999996$. Just this tiny shift in $\lambda$ raises $n_s$ sufficiently that the model agrees with data for the full ranges of possible reheating temperatures. Values of $\lambda > 1$ further push $n_s$ to lower values.

\begin{figure}[ht!]
\includegraphics[width=.45\textwidth]{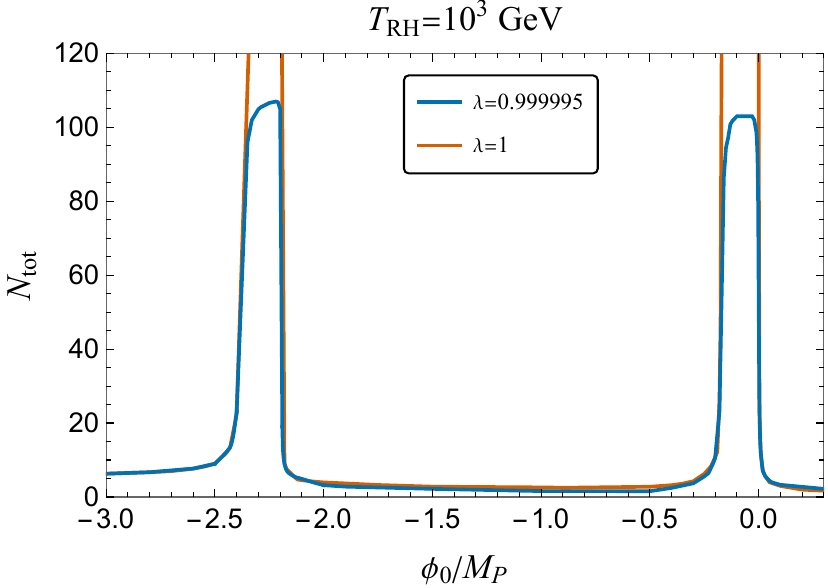}\hspace{10mm}
\includegraphics[width=.45\textwidth]{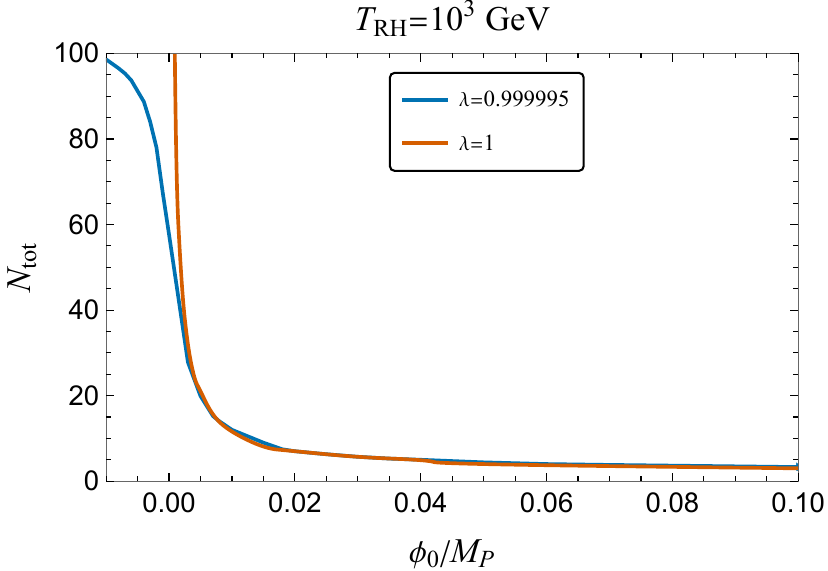}
\caption{\it Left panel: the total number of $e$-folds, $N_{\rm tot}$, as a function of $\phi_0$ for $\lambda = 1$ and  $\lambda = 0.999995$ with $K=\phi^* \phi$ and superpotential $W=M (\lambda  - \phi )^2$. Right panel: zoom of the same plot in the neighbourhood of $\phi_0 = 0$.}
\label{noteternal2}
\end{figure}

\begin{figure}[ht!]
\includegraphics[width=\textwidth]{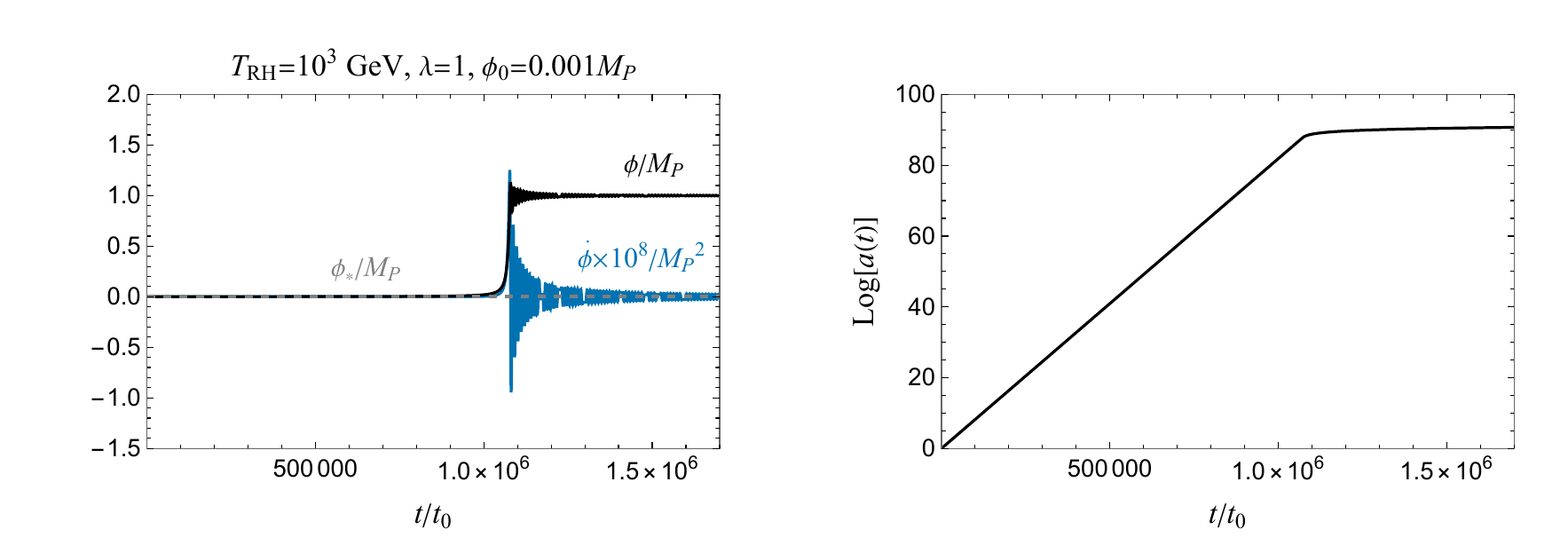}\\
\includegraphics[width=\textwidth]{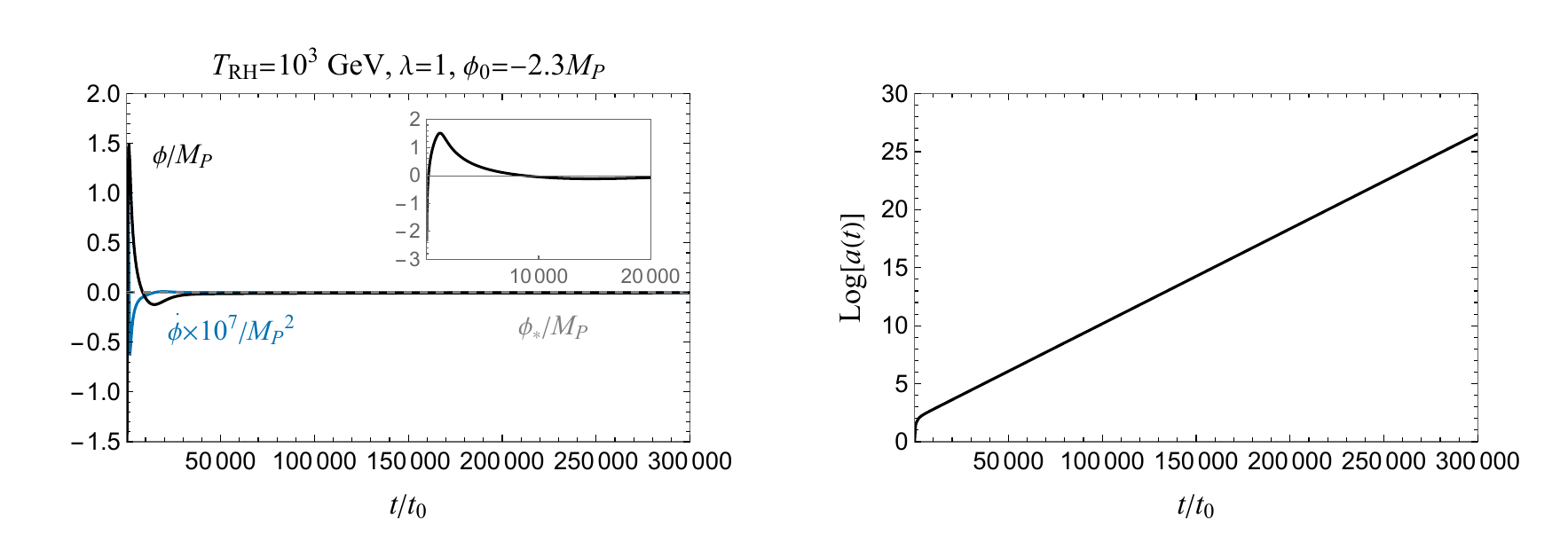}\\\includegraphics[width=\textwidth]{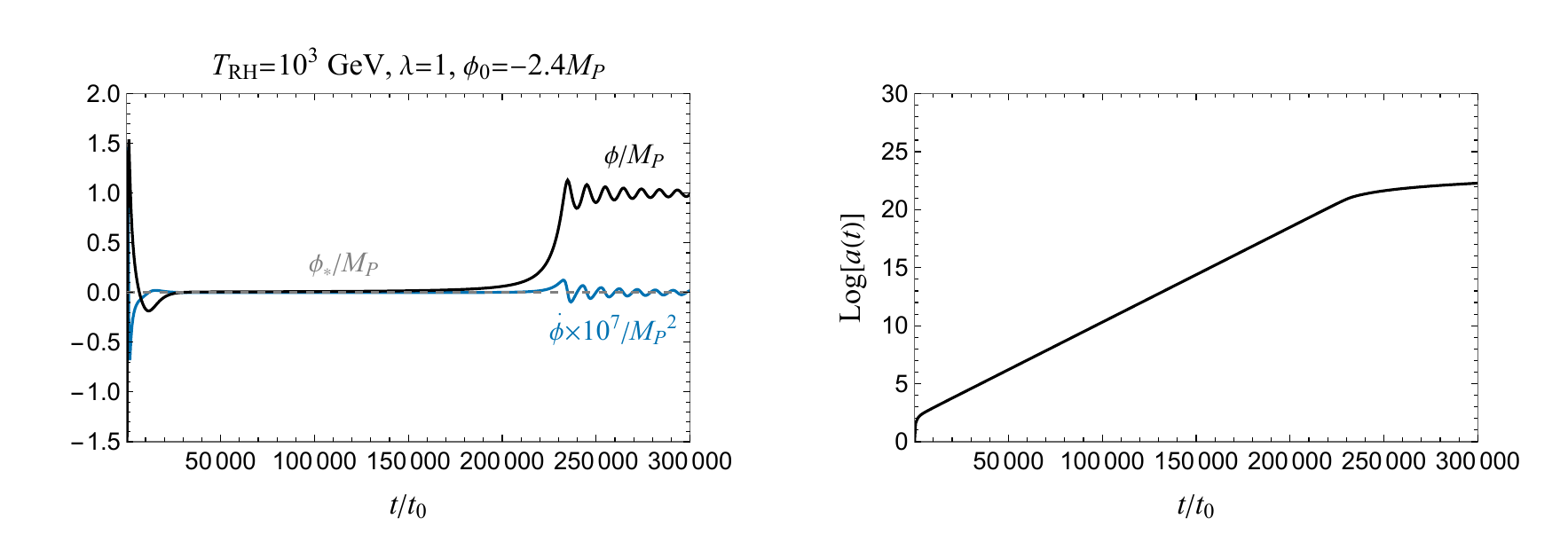}
\caption{\it The evolution of the inflaton as a function of time (left) and the total number of $e$-folds of inflation (right) for $\lambda=1$ in the HRR model and different choices of $\phi_0$. At $\phi_0=-2.3 M_P$ (middle panels), the inflaton reaches $\phi\simeq 0$ after one period of oscillation and then stays frozen.}
\label{newinffrozen}
\end{figure}

There are other interesting aspects to the HRR model. Normally, one assumes that $\phi_0 \approx 0$ to get inflation\footnote{One reason why this model was initially disfavored was the belief that the initial condition should be set by the minimum in the finite-temperature effective potential~\cite{gno,ovs,os,Binetruy:1984wy}. In the HRR model, the finite-temperature minimum is not near $\phi = 0$, and this was thought to preclude the possibility of slow-roll inflation, motivating the construction of a similar no-scale version of new inflation \cite{Ellis:1984bf,Enqvist:1984jj}. However, in the absence of a pre-inflationary thermal bath these corrections are not important, and the initial conditions can be taken as arbitrary, though field values that lead to more inflationary $e$-folds will dominate the physical volume and can be interpreted as more probable initial values.}. In Fig.~\ref{noteternal2}, we show the total number of $e$-folds obtained as a function of $\phi_0$ for $\lambda = 1$ and 0.999995,  $\trh = 10^3$~GeV, assuming ${\dot\phi}_0 = 0$.  There are two distinct peaks in $N_{\rm tot}$. For $\lambda = 1$ (shown by the red line), if $\phi_0 = 0$, $N_{\rm tot}$ diverges classically, as $\dot\phi$ remains 0, and the inflaton does not evolve. Then, for slightly larger values of $\phi_0$ there are a finite number of $e$-folds, as seen more clearly in the right panel of  Fig.~\ref{noteternal2}, which zooms in on the region near $\phi = 0$. For example, for $\phi_0 = 0.0015 M_P$ we have $N_{\rm tot} \simeq 55$. The evolutions of $\phi$ and $\dot\phi$ are shown for $\phi_0 = 0.001$ in the upper panels of Fig.~\ref{newinffrozen}. After a sufficient period of inflation the inflaton starts oscillations about its minimum.  In contrast, $N_{\rm tot} \sim 9$ for $\phi_0 = 0.01 M_P$. For $\lambda < 1$, there is no longer an inflection point in the potential and the total number of $e$-folds is limited to $\sim 100$ for $\lambda = 0.999995$ (shown by the black curve). In either case ( $\lambda = 1$ or $\lambda = 0.999995$), it is also possible to take $\phi \simeq -0.3 M_P$ and obtain a sufficient amount of inflation. In this case, $\dot\phi$ is small though non-zero 
as the inflaton passes through the origin. Thus, near $\phi = 0$ there are two initial conditions with $N_* \sim 50$. For smaller (more negative) $\phi_0$, 
the field starts moving too quickly 
and reaches the minimum of the potential before any substantial expansion occurs.

As also seen in the left panel of Fig.~\ref{noteternal2}, there is a second peak in $N_{\rm tot}$ at $\sim -2.3M_P$ corresponding to a point where the potential is very steep, and
the classical evolution of $\phi$ carries the inflaton past its minimum and back to the inflection point where inflation can commence.
At $\phi_0 \approx -2.3 M_P$,
the field asymptotically resettles at $\phi = 0$ and stops with inflation never ending (of course quantum fluctuations will spoil this naive picture). The evolution of this special case is shown in the middle panels of Fig.~\ref{newinffrozen}. The inflaton quickly evolves to $\phi >1$ and resettles eternally at $\phi = 0$ and continues to expand exponentially. At still lower values of $\phi_0 < -2.3 M_P$,
the inflaton evolves past its minimum and returns to an initial 
value where inflation occurs as if $\phi_0 \sim - 0.3 M_P$. This is seen in the lower panels of Fig.~\ref{noteternal2} where $\phi_0 = -2.4 M_P$.  In principle there are an infinite number of such peaks (at both lower and higher values of $\phi_0$, but all other peaks require large values of $|\phi|$ such that  $V(\phi_0) > M_P^4$. These initial conditions are widely considered to be unphysical, and are not considered here.

\begin{figure}[ht!]\begin{center}
\includegraphics[width=.75\textwidth]{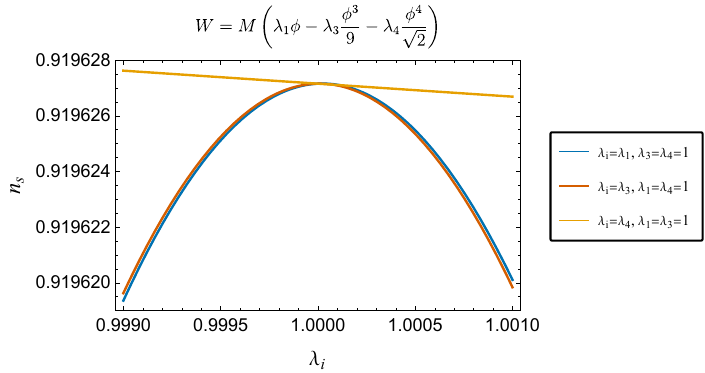}
\end{center}
\caption{\it Values of $n_s$ as functions of $\lambda_1,\lambda_3,\lambda_4$ in the model defined by Eq.~(\ref{eenos}). Here we fix $\trh=10^{10}\gev$. }
\label{nsvl1eenos}
\end{figure}

As noted above, a similar class of new inflation models can be constructed in no-scale supergravity \cite{Ellis:1984bf,Choi:2024ruu}. The initial motivation (now considered moot, as discussed above) for these was to restore the high-temperature minimum setting the initial conditions for $\phi$. The model defined by
\beq
W = M \left(\lambda_1 \chi - \lambda_3 \frac{\chi^3}{9}
- \lambda_4 \frac{\chi^4}{\sqrt{2}} \right)
\label{eenos}
\eeq
yields a potential very similar to the HRR model when $\lambda_i = 1$. However, this model
also gives $n_s \simeq 0.92$ as in the new inflation model shown in Fig.~\ref{nsvtrhrr}, and in this case varying any one of the three $\lambda_i$ drives $n_s$ to lower values as shown in Fig.~\ref{nsvl1eenos},
so the model cannot be saved by varying $\lambda_i$~\footnote{However, it can be saved with the help of a curvaton \cite{Choi:2024ruu}.}. We do not discuss this model any further.

\section{Swampland Distance Conjecture}
\label{swamp}

As mentioned in the Introduction, one of the issues with Starobinsky inflation (and all large-field inflation models) is the need for large initial field values. If  the inflaton and therefore its displacement take super-Planckian values,  the validity of the effective field theory is generally broken. The breaking is quantified by the Swampland Distance Conjecture, which implies the appearance of a tower of `light' states with masses that are exponentially small in the proper distance with an exponent of order unity in four-dimensional Planck units~\cite{Ooguri:2006in}. This tower is in general connected to the decompactification of extra dimensions or to a string theory tower of states~\cite{Lee:2019wij}, which also implies that the `light' particles include massive spin-2 states. These give rise to problems with unitarity if  they are lighter than the Hubble constant during inflation~\cite{Higuchi:1986py}. The scale of the tower should therefore be larger than the inflation scale.

To be more precise, the Swampland Distance Conjecture can be formulated as a limit on the trans-Planckian excursions of the scalar fields in the theory:
\beq
\Delta \phi < c \ln \frac{M_P}{Q}\, ,
\eeq
where $c$ is some constant of order 1, and $Q$ is the tower scale and thus the highest scale for which an effective field theory approach without extra states is appropriate. 
The inflationary scale is considerably below the Planck scale, so taking $Q\sim H$ would indicate that field excursions should be no larger than $\mathcal{O}(10)$. The precise details of the breakdown are model-dependent, and we use the string-derived ANR model as a guide.~\footnote{For further discussion of the Swampland Conjecture in the context of no-scale supergravity and the Starobinsky model see \cite{Scalisi:2018eaz, Rasulian:2021wny, Lust:2023zql,ano}.}

The superpotential in the ANR model \eqref{Wstring} can be brought in the same form as in Eq.~(\ref{wii}), by performing the transformation
\beq
\label{ytoT}
y=\frac{2T-1}{2T+1}\,,
\eeq
leading to
\beq
\label{TzBasis}
K_\text{string}=-2\ln(T+{\bar T})-2\ln\left(1-\frac12|z|^2\right)\quad;\quad W=M z\left(T-\frac12\right)\,.
\eeq
Inspection of the above K\"ahler potential in the context of the compactification of ten-dimensional supergravity on a $\mathbb{Z}_2\times\mathbb{Z}_2$ orbifold allows one to identify $T$ with the area modulus of a torus $T^2$, so that $2T=R^2+ib$ in string units with $R^2$ being the area of $T^2$ and $b$ the corresponding axion \cite{ano}. Their kinetic terms are then consistent with those obtained via dimensional reduction of the Einstein-Hilbert action in $4+d$ dimensions compactified on a $d$-dimensional square torus of radius $R$:
\beq
\label{dimredd}
\frac12{\cal R}^{(4+d)} \longrightarrow\frac12{\cal R}^{(4)}-\frac{d(d+2)}{4}\left(\frac{\partial R}{ R}\right)^2 \, ,
\eeq
for $d=2$. The compactification scale associated with the light tower of states is $Q \sim g_s (R)^{-(2+d)/2}$ in string units, where $g_s \sim 0.2$ is the string coupling. Thus, for $d=2$ we have $Q \sim g_s e^{-2\phi}$, which allows $\phi \sim 10 M_P$ for $Q\sim H$.

The above analysis can be repeated for the $R+R^2$ supergravity model with K\"ahler potential \eqref{v0} and superpotential \eqref{chargestaro}\footnote{Recall that the superpotentials (\ref{chargestaro}) and (\ref{wii}) are equivalent up to a field redefinition and K\"ahler transformation. Similarly, the superpotentials (\ref{wi}) and (\ref{wsb1}) are equivalent. The two classes are also related by a SU(2,1)/SU(2)$\times$U(1) transformation \cite{enov1}.  Here we refer to them collectively as SUSY $R+R^2$.}. In this case, $T$ is the area modulus of a 4-cycle with $2 T=R^4+ib$ in string units and all mutually orthogonal 4-cycles having the same size. Agreement with the dimensionally-reduced action \eqref{dimredd} implies that $d=6$, consistent with the argument above. From Eq.~(\ref{canphi}) we see that the canonically-normalized inflaton is $\phi = 2\sqrt{6} \ln R$.
The compactification scale associated with the light tower of states is then $Q \sim g_s (R)^{-(2+d)/2}$ in string units, where $g_s \sim 0.2$ is the string coupling. Thus, for $d=6$ we have $Q \sim g_s e^{-\sqrt{2/3}\phi}$, which allows $\phi \sim 12 M_P$ for $Q\sim H$.

\begin{figure}[ht!]\begin{center}
\includegraphics[width=.55\textwidth]{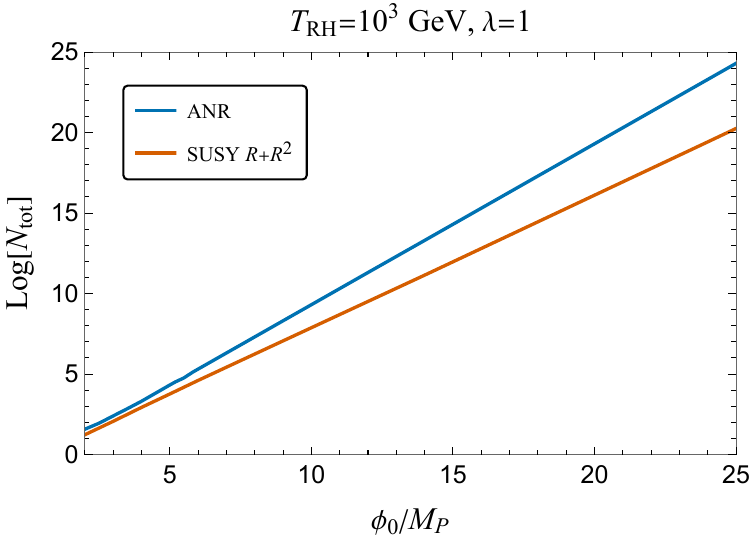}
\caption{\it The total numbers of $e$-folds, $N_{\rm tot}$, as functions of $\phi_0$ for $\lambda = 1$  in the   SUSY $R+R^2$ and ANR models.}
\label{eternal}
\end{center}
\end{figure}

If a measure of the probability that we come from a patch of the Universe is related to the total number of $e$-folds produced, we would more likely come from extremely large initial values of $\phi$, in violation of the large-distance conjecture. This can be seen in Fig.~\ref{eternal},
which shows that the total number of $e$-folds, $N_{\rm tot}$, rises exponentially with $\phi_0$ for the Starobinsky model $(\lambda = 1)$. In other words, patches of the Universe originating from large initial field values have exponentially larger volumes when inflation is complete, and the probability that we came from such a patch is exponentially larger than the probability of coming from a patch with a small initial field value. 

However, when $\lambda < 1$ in the SUSY $R+R^2$ inflationary model~\cite{eno6, Cecotti} or in the ANR model~\cite{anr1}, there is a barrier preventing the initial values from taking arbitrarily large values, enabling the swampland problem to be avoided.
Successful inflation with acceptable $N_{\rm tot}$, $n_s$, and $\trh$
requires $\lambda > 0.99984$ (0.999928) in the SUSY $R+R^2$ (ANR) model. For the Planck + ACT + DESI BAO combination of data, this limit is $\lambda > 0.99979$ (0.99984).

For these values of $\lambda$, $V(\phi_0) < M_P^4$ for $\phi< 25.1 M_P$, which is deep in the swampland since, as argued above, one needs $\phi_0 \lesssim 12 M_P$ to avoid swampland issues. 

\begin{figure}[ht!]
\includegraphics[width=.5\textwidth]{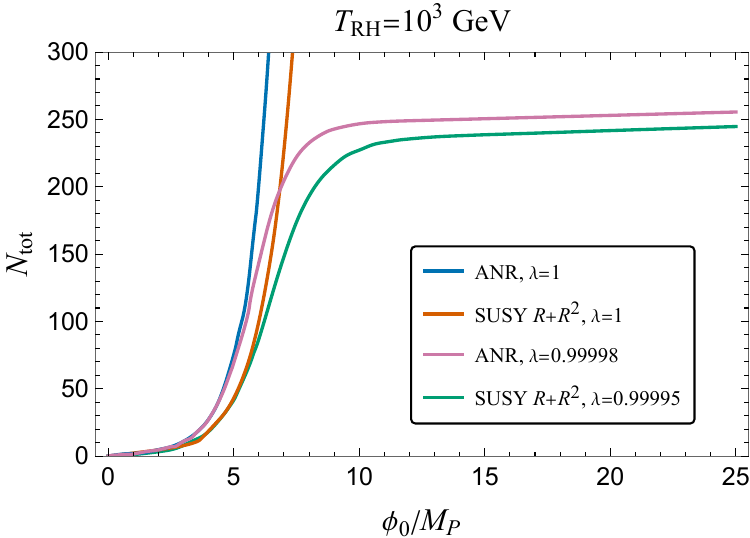}\includegraphics[width=.5\textwidth]{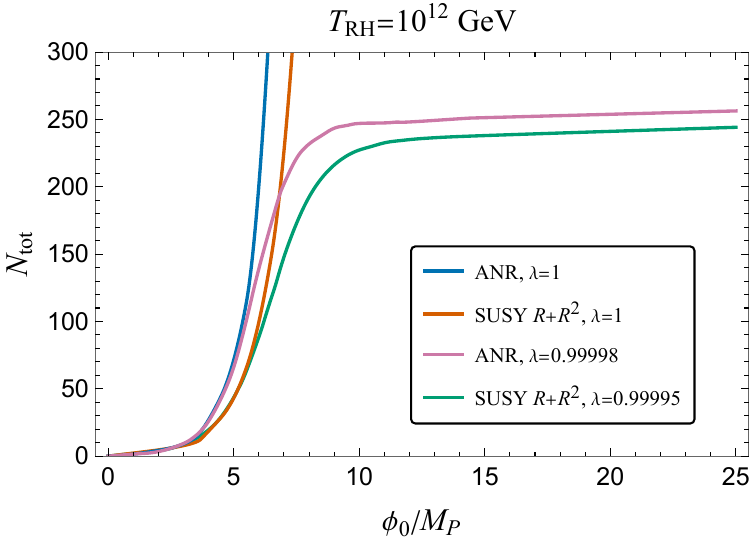}
\caption{\it The total numbers of $e$-folds, $N_{\rm tot}$, as functions of $\phi_0$ for different values of $\lambda $ in the  SUSY $R+R^2$ and ANR models.}
\label{lam1and9}
\end{figure}

However, the total number of $e$-folds in the models with $\lambda < 1$ is not very sensitive to $\phi_0$. We show in Fig.~\ref{lam1and9} 
for $\trh = 10^3$~GeV and $10^{12}$~GeV the total numbers of $e$-folds as functions of $\phi_0$.
Also shown in these figures is the comparison to the Starobinsky model (with $\lambda = 1$). 
As one can see, once
$\phi_0 \gtrsim 8 M_P$, the total number of $e$-folds varies little. For large $\phi_0$, the inflaton evolves very quickly before settling to a slow roll as was depicted in Fig.~\ref{vlam1}.  This implies a relatively flat probability distribution for $\phi_0$ over the range $8 - 25 M_P$. Thus the probability that our patch of the Universe originates from $\phi_0 = 8 - 12 M_P$
is not significantly less than the probability that it originates from $\phi_0 > 12 M_P$, thus potentially providing a solution to the swampland problem for Starobinsky-like models. 

 Finally, we recall that although there is no swampland issue related to the HRR model (or any small field inflation model), the choice of $\lambda < 1$ was necessary in order to boost the value of $n_s$ to acceptable values.

\section{Conclusions}
\label{summ}

Any model of inflation must be concordant with experimental observables, notably $n_s$ and $r$. At the same time, the model should be well grounded in theory. 

All of the examples discussed in Section \ref{examples}, whether Starobinsky-like or new inflation models of inflation, are formulated as ``accidental" models,
in the sense that some coupling parameter $\lambda$ must be specified to high precision. There is no conceptual distinction between a theory with $\lambda = 0.99995$ or one with $\lambda = 1.00000$: they are just (slightly) different accidents. Nevertheless, as we have shown, a slightly different accident can occasionally save a model both experimentally and theoretically.

The Starobinsky model of inflation in its original geometrical formulation has no clear relation to string theory. Moreover, whether in its original formulation or when derived from no-scale supergravity or string theory with $\lambda = 1$, it  necessarily involves large field excursions. Indeed, regions with large initial field values, $\phi_0$, expand the most (see Fig.~\ref{eternal}) making it more likely that our universe originated from such a region.

On the other hand, as well as having foundations in underlying theoretical ideas, both the no-scale and ANR string formulations of Starobinsky-like models offer simple possibilities for modifying the potential. 
In these models, the Starobinsky potential is obtained by a unique (accidental) choice of couplings, $\lambda =1$.
However, when $\lambda <1$ the infinite plateau at large field values appearing in the Starobinsky model is lifted (see Fig.~\ref{eno6lambda}), with a potential barrier to reaching very large field values, enabling the models to be consistent with the Swampland Conjecture. 
Actually, in string theory the value of the $\lambda$ is calculable in principle. In particular, in the ANR model $\lambda\sim{\cal O}(1)$ is not accidental as was explained in Section \ref{ANR}.

The Starobinsky model is in good agreement with the Planck determination of $n_s$~\cite{planck18}, as are the no-scale and ANR models with $\lambda = 1$. However, a slightly higher value of $n_s$, as perhaps indicated by the combination of Planck and ACT data or the combination of Planck + ACT + DESI BAO data~\cite{ACT:2025fju}, is in tension with the Starobinsky model. The combined data are easier to accommodate in the no-scale and ANR models with $\lambda < 1$, which is not possible in the original geometrical formulation of the Starobinsky model. 

As well as accommodating a higher value of $n_s$, the potential barrier appearing in the no-scale and ANR models when $\lambda < 1$ can keep the model out of the swampland~\cite{Lust:2023zql} by limiting the initial field value to $\phi_0 \lesssim 12 M_P$ so that the mass scale associated with the infinite tower of light states predicted in string theory remains above the Hubble scale. In addition, we have shown that the total number of $e$-folds is limited to $\lesssim 250$ in the models with $\lambda < 1$ (see Fig.~\ref{lam1and9}), almost independent of $\phi_0$ for $\phi_0 \gtrsim 8$. Thus the likelihood that our patch of the universe originated in a non-swampy scenario with $\phi_0 < 12 M_P$ is no less likely than having originated in a swampy scenario with $\phi_0 \gtrsim 12 M_P$. 

For context, we have also discussed the HRR new inflation model \cite{hrr}, which was derived in the context of minimal supergravity  with a superpotential of the form $(1-\phi)^2$, corresponding to another ``accidental” choice $\lambda = 1$. As a small-field inflation model, there are no issues regarding the Swampland Conjecture, since field excursions are always of order $M_P$. On the other hand, the HRR model predicts $n_s \simeq 0.92$ over a wide range of reheating temperatures, in conflict with the CMB data. However, the scalar potential in the HRR model is also very sensitive to $\lambda$. Concentrating on values of $\lambda \le 1$ to avoid the appearance of additional minima, we have found that for certain values of $\lambda < 1$ the potential is sufficiently altered that $n_s$ can be brought into agreement with either Planck or the combination of Planck, ACT and DESI BAO data (see Fig.~\ref{vlamhrr}).~\footnote{When $\lambda > 1$ the value of $n_s$ is even more discrepant with experiment.} The tensor-to-scalar ratio is largely independent of $\lambda$ and is always far below the current experimental limits.

Experiment will be the ultimate arbiter that judges the viability of models of inflation. So far, models whose predictions resemble those of the original Starobinsky model are consistent with the data, though it may be in tension with ACT and DESI BAO data. Supergravity-based models can reproduce the Starobinsky predictions for specific ``accidental" values of key model parameters, but small deviations from these values accommodate the ACT and DESI BAO data more readily. Moreover, these slightly different ``accidental" choices predict the existence of potential barriers that prevent the supergravity models from straying into the swampland at large field values. Time will tell whether this double preference for ``accidental" no-scale supergravity-based models will be confirmed.

\acknowledgments
I.A. is supported by the Second Century Fund (C2F), Chulalongkorn University. The work of J.E. was supported by the United Kingdom STFC ST/T000759/1. The work of K.A.O.~was supported in part by DOE grant DE-SC0011842 at the University of Minnesota.

\end{document}